\documentclass[
 twocolumn
,secnumarabic
,amssymb 
,amsmath 
,nobibnotes
,aps
,superscriptaddress]{revtex4}

\usepackage{bm}
\usepackage{graphicx}

\usepackage[colorlinks=false,linkcolor=blue]{hyperref}
\expandafter\ifx\csname package@font\endcsname\relax\else
 \expandafter\expandafter
 \expandafter\usepackage
 \expandafter\expandafter
 \expandafter{\csname package@font\endcsname}%
\fi

\begin{document}

\title{Random matrix ensembles of time-lagged correlation matrices: Derivation of  
       eigenvalue spectra and analysis of financial time-series}

\author{Christoly Biely}
\email{christoly.biely@meduniwien.ac.at}
\author{Stefan Thurner}
\email{thurner@univie.ac.at}
\affiliation{Complex Systems Research Group, HNO, Medical University of 
Vienna,
W\"ahringer G\"urtel 18-20, A-1090 Vienna 
\\ and \\
Atominstitut der \"Osterreichischen Universit\"aten, Stadionallee 2, A-1020 Vienna, Austria}

\begin{abstract}
We derive the exact form of the eigenvalue spectra of correlation matrices derived from a set of time-shifted, 
finite Brownian random walks (time-series). These matrices can be seen as random, real, asymmetric
matrices with a special structure superimposed due to the time-shift. 
We demonstrate that the associated eigenvalue spectrum is  circular symmetric in the complex plane
for large matrices.  This fact allows us to exactly compute the 
eigenvalue density  via an inverse Abel-transform of the density of the {\it symmetrized} problem.
We demonstrate the validity of this approach by numerically computing  
eigenvalue spectra of lagged correlation matrices based on uncorrelated, Gaussian distributed time-series.
We then compare our theoretical findings with eigenvalue densities obtained from actual high frequency (5 min) data 
of the S\&P500 and discuss the observed deviations. 
We identify various non-trivial, non-random patterns and find asymmetric dependencies associated with 
eigenvalues departing strongly from the Gaussian prediction in the imaginary part.
For the same time-series, with the market contribution removed, we observe strong clustering of stocks, 
i.e.  {\it causal sectors}. We finally comment on the time-stability of the observed patterns.\\


\end{abstract}

\date{\today}

\maketitle

\section{Introduction}

One of the pillars of contemporary theory of financial economics is the 
notion of correlation matrices of timeseries of financial instruments;
the capital asset pricing  model \cite{sharpe} and Markowitz portfolio 
theory \cite{markowitz} 
probably being the most prominent examples. 
Recent empirical analyses on 
the detailed structure
of financial correlation matrices have shown that there exist remarkable 
deviations from predictions that would be expected from the efficient market 
hypothesis. In particular, based on pioneering work \cite{PRLs1,PRLs2}, 
eigenvalue spectra of empirical equal-time covariance matrices have been analyzed and  
compared  to predictions of eigenvalue densities for Gaussian-randomness obtained from
random matrix theory (RMT).
It has been shown, that the eigenvectors which strongly depart from the spectrum
obtained by RMT contain information about sector organization of 
markets \cite{plerou,systematic}. The largest
eigenvalue has been identified as the 'market-mode', and it has been 
pointed out that
a 'cleaning' of the original correlation matrices by removing
the noise part of the spectrum explainable by RMT results in an improved 
mean variance efficient frontier which seems to be much more adequate than the one obtained by 
Markowitz (see e.g. the recent discussion in \cite{oldlaces}). 
Further, RMT provides an almost full understanding of why the Markovitz approach 
is close to useless (dominance of small eigenvalues which lie in the noise regime) in actual portfolio management.  

Initially, RMT has been proposed to explain energy spectra of 
complicated nuclei half a century ago.
In its simplest form, 
a random matrix ensemble is an ensemble of $N\times{}N$ matrices ${\bf M}$ whose entries 
$M_{ij}$ are uncorrelated iid random variables, and whose 
distribution is given by
\begin{equation}
P({\bf M}) \sim \exp \left(- \frac {\beta N}{2} \mathrm{Tr}({\bf MM}^{T}) \right) \quad ,
\end{equation}
where $\beta$ takes specific values for different 
ensembles of matrices (e.g. depending on whether or not the random variables are complex- or real-valued). 
Eigenvalue spectra and correlations of eigenvalues in the limit
$N\rightarrow\infty$ have been worked 
out for {\em symmetric} 
$N\times{}N$
random matrices by Wigner \cite{wigner}. 
For real valued matrix entries, such symmetric random matrices are sometimes referred to as the 
Gaussian orthogonal ensemble (GOE). 

The symmetry constraint has later been relaxed by Ginibre 
and the probability distributions of different ensembles
(real, complex, quaternion)  
-- known as Ginibre ensembles (GinOE, GinUE, GinSE) --
have been derived \cite{ginibre} in the limit of infinite matrix size.
For ensembles of random real asymmetric matrices (GinOE) -- the most difficult case -- 
progress has only slowly been made under great efforts over the past decades.   
The eigenvalue density could finally be derived via different methods
\cite{crisantisommers2, edelman}, where -- quite remarkably -- the 
finite-size dependence of the ensemble has also been elucidated \cite{edelman}. 
For recent progress in the field also see \cite{kanzieper}.

However, these developments in  RMT do not yet take into account the timeseries 
character of financial applications, i.e. the fact, that one deals -- in general -- with 
(lagged) covariance  matrices stemming from finite rectangular $N \times T$ data matrices
${\bf X}$, which contain 
data for $N$ different assets 
(or instruments) at $T$ observation points. 
The matrix ensemble corresponding to the $N\times{}N$ covariance matrix ${\bf C} \sim {\bf X}{\bf X}^T$
of such data is known as the  Wishart ensemble \cite{wishart} and is a cornerstone of
multivariate data analysis. For the case of uncorrelated
Gaussian distributed data, the exact solution to the eigenvalue-spectrum of
${\bf X}{\bf X}^T$ is known as Marcenko-Pastur law (for $N\rightarrow\infty$) and has been used as a 
starting point for random matrix analysis of correlation matrices at lag zero
\cite{PRLs1,PRLs2,plerou,systematic,oldlaces,nocheiner}. 
Moreover a quite general methodology of extracting meaningful correlations between
variables has been discussed based on a generalization of the Marcenko-Pastur
distribution \cite{Bouchaud}. The underlying method was the powerful tool of singular-value decomposition
and RMT was used to predict singular-value spectra of Gaussian randomness.

The time-lagged analogon to the covariance matrix is defined as 
$C_{\tau}^{ij}\sim  \sum_t^T  r^i_t r^j_{t-\tau}$,
where one timeseries is shifted by $\tau$ timesteps with respect to the other.
In contrast to (real-valued) equal-time correlation matrices of the Wishart ensemble, which 
have a real eigenvalue spectrum, the spectrum of ${\bf C}_{\tau}$  is defined
in the complex plane since matrices of these type are in general asymmetric.
While the complex spectrum of ${\bf C}_{\tau}$ remains unknown so far, 
results for \emph{symmetrized} lagged correlation matrices 
have been reported recently \cite{delay_corr,burda_financial}.
In \cite{burda_financial}, it was also shown that the methodology 
of free random variables can be used to tackle a variety of correlated (symmetric) Wishart 
matrix models.

However, it is the analysis of the initial asymmetric time-lagged correlations 
which forms a fundamental part of finance and econometrics, 
and which  has attracted considerable attention in the respective literature. 
The existence of asymmetric lead-lag relationships has been initially reported
for the U.S. stock market \cite{lo_kinlay}. Specifically, it was found that returns of large stocks lead those
of smaller ones. Later, trading volume was identified as  a significant
determinant of such lead-lag patterns, and returns of high-volume stocks 
(portfolios) were found to lead those of low-volume stocks (portfolios) \cite{chordia}. 
These lead-lag effects have primarily been explained by different effects of information
adjustment asymmetry. For instance, a model was brought forward in \cite{chan},
where it was argued,
that, as soon as previous price changes are observed and marketwide information can
thus be incorporated in the marketmakers' evaluation of stock prices, 
lagged correlations may emanate. 
Another type of information asymmetry can be seen in the different number of investment
analysts following a firm's stock price \cite{brennan}.
Other explanatory approaches,  
include the institutional ownership of
stocks \cite{badrinath},
the different exposure of stocks to persistent factors \cite{hameed}, 
or transaction costs and market microstructure \cite{mench} as causes of
lagged autocorrelations. 
Whether or not
non-synchronous trading may constitute a source of lead-lag relationships or
not is an issue of ongoing discussion \cite{lo_kinlay,boudoukh,bernhardt}. 
Recently, aiming at a closer empirical understanding of lagged correlations, 
the dependence of the strength of lagged correlations on the chosen
time-shift $\tau$ has been analyzed  
for high-frequency  NYSE data \cite{kertesz2}. It was shown, that the lagged correlation 
function typically exhibits an asymmetric peak. The revealed patterns  basically showed structures
consistent with those found in \cite{lo_kinlay} 
(e.g. patterns where more 'important' companies pull smaller, less 'important' ones).
Interestingly, also evidence for a diminution of the Epps effect \cite{epps} has been
demonstrated based on lagged cross-correlations of NYSE-data, as lead-lag dependencies
seem to diminish over the years \cite{kertesz1}.

As diverse, interesting and as on-going these approaches are, 
the methods applied are mainly based on Granger causality, 
vector autoregressive models and shrinkage estimators.
In this paper, we want to extend the methodology to eigenvalue analysis of
time-lagged correlations.
First, we discuss how solutions of RMT problems pertaining to real, asymmetric
matrices can be obtained from solutions to the symmetrized problem via an
inverse Abel-transform.
The respective developments will then enable us to 
derive the form of the eigenvalue spectra of the pure random case. 
As an immediate application we compare these theoretical results,  
with real financial data and relate the observed deviations 
to market specific features.  

The paper is organized as follows: 
In Section \ref{sec2} we fix the notation and develop the spectral form of asymmetric 
real random correlation matrices. 
In Section \ref{sec3} we apply the introduced methodology to empirical 
correlation matrices of 5 min log-returns of the S\&P500 and discuss 
the meaning of deviant eigenvalues from several perspectives.
Time-dependence issues are discussed  
in Section \ref{sec4} and  
in Section \ref{sec5} we finally conclude.

\section{Spectra of time-lagged correlation matrices}
\label{sec2}

\subsection{Notation}

The entries in the $N\times{}T$ data matrices ${\bf X}$ for $N$ assets and $T$ observation times, 
are the log-return time-series of asset $i$ at observation times $t$, 
\begin{equation}
r_{t}^{i}=\ln{}S_{t}^i-\ln{S_{t-1}^i} \quad ,
\end{equation}
after subtraction of the mean and normalization to unit variance, i.e. division by 
$\sigma_i = \sqrt{\langle{}(r^i_{t})^2\rangle-\langle{}r^i_{t}\rangle^2}$. 
Here, $S_t^i$ is the price of asset $i$ at time $t$.
One time unit is the time difference between observations at $t+1$ and $t$, 
e.g. a day, 5 minutes; for tic data it can also be of variable size.
Time-lagged correlation functions of unit-variance log-return series among stocks 
are defined as
\begin{equation}
  C^{ij}_{\tau} (T) \equiv 
  {\langle{}(r^i_{t}-\langle{}r^i_{t} \rangle) (r^j_{t-\tau}-\langle{}r^j_{t-\tau} \rangle)\rangle}_T
  \quad , 
\end{equation}
where the time-lag $\tau$ is  measured in time units 
and $\langle{}...\rangle{}_T$ stands for a time-average over the period $T$.
We drop $(T)$ in the following, except for Section \ref{sec4}. 
Equal-time correlations are obviously  obtained for $\tau=0$.
For $\tau\not=0$, the lagged correlation matrix $\mathbf{C}_{\tau}$
is generally not symmetric and contains the lagged 
autocorrelations in the diagonal. It can be written as 
\begin{equation}
\label{Cdefinition}
  {\bf C}_{\tau}=\frac{1}{T}{\bf X}{\bf D}_{\tau}{\bf X^T}\quad,  
\end{equation}
where ${\bf D}_{\tau} \equiv \delta_{t,t+\tau}$ and where ${\bf X}$ is the $N\times{}T$ 
normalized time-series data.
Denoting the eigenvalues of $C^{ij}_{\tau}$ by $\lambda_i$ and their associated
eigenvectors by $\vec{u}_i$ (or $u_{ik}$), where $i,k=1,...,N$, we may write the 
eigenvalue problem as 
\begin{equation}
\label{eigenvalueproblem}
\sum_{j}C^{ij}_{\tau}\vec{u}_j=\lambda_j\vec{u}_{j} \quad .
\end{equation}
We immediately recognize that eigenvalues $\lambda_i$ are either real or complex
conjugate, since the matrix elements of $C^{ij}_{\tau}$ are real and thus the
conjugate eigenvalue $\lambda_i^*$ also solves Eq. (\ref{eigenvalueproblem}).
Regarding the elements of $C^{ij}_{\tau}$ as random variables with a
certain distribution, we should keep in mind that their specific construction,
Eq. (\ref{Cdefinition}), results in a departure from a 'purely' random real
asymmetric $N\times{}N$ matrix where the entries are iid Gaussian distributed.
Thus we do -- in general --  \emph{not} expect a flat eigenvalue distribution as
in the Ginibre-Girko case.
Rather, we can interpret ${\bf C}_\tau$ as a random real asymmetric matrix
with a special structure due to its construction.
In general, comparably little work has been done to understand the eigenvalue spectra of
such random real asymmetric matrices. 
Unfortunately, powerful addition formalisms developed for non-Hermitian random
matrices (see e.g. \cite{pappzahed1} and references therein) 
are not applicable in the case of random real asymmetric matrices. 
However, it was
shown that the problem can be treated in a way formally equivalent to
classical electrostatics \cite{crisantisommers,crisantisommers2} and a
generalization of  Girko's semicircular law \cite{girko} could be recovered
via application of the replica-technique. 

\subsection{General Arguments}

We start our arguments from the electrostatic potential analogy, originally
introduced by Wigner.
The idea is to interpret the distribution of eigenvalues in the complex plane as 
a distribution of electrical charges in 2 dimensions.  
Following the same arguments as in \cite{crisantisommers}, the corresponding potential in 2 dimensions is given by 
\begin{equation}
\label{potential}
 \phi(x,y)=-\frac{1}{N}\langle{}\ln\det\left((\delta_{ij}z^*-C_{\tau}^{ji})(\delta_{ij}z-C_{\tau}^{ij})\right)\rangle{}_c
  \quad, 
\end{equation}
where $z=x+iy$, and $\langle...\rangle{}_c$ denotes the average over the distribution, 
\begin{equation}
\label{dist}
 P({\bf X}) \sim{} \exp\left(-\frac{N}{2}\mathrm{Tr}({\bf X}{\bf ^T})\right) \quad , 
\end{equation}
of the matrices $X_{ij}$. 
It can be shown \cite{crisantisommers} that Eq. (\ref{potential}) allows for the calculation of a density 
$\rho(z)=\rho(x,y)$ via the Poisson equation
\begin{equation}
  \rho(x,y)=-\frac{1}{4\pi}\triangle{}\phi(x,y) \quad. 
  \label{rho2}
\end{equation}

Expanding the argument of the determinant in Eq. (\ref{potential}) we obtain
the positive definite matrix
\begin{equation}
\label{argument}
H_{ij}=\delta_{ij}|z|+C_{\tau}^{ij}C_{\tau}^{ji}-x(C_{\tau}^{ij}+C_{\tau}^{ji})+iy(C_{\tau}^{ij}-C_{\tau}^{ji})
\quad .
\end{equation}
This form shows that any symmetric (anti-symmetric) contribution of $C_{\tau}^{ij}$
only influences the real (imaginary) part of $z$. 

\begin{center}
\begin{figure}[h!]
\begin{center}
\begin{tabular}{cc}
\resizebox{0.25\textwidth}{!}{\includegraphics{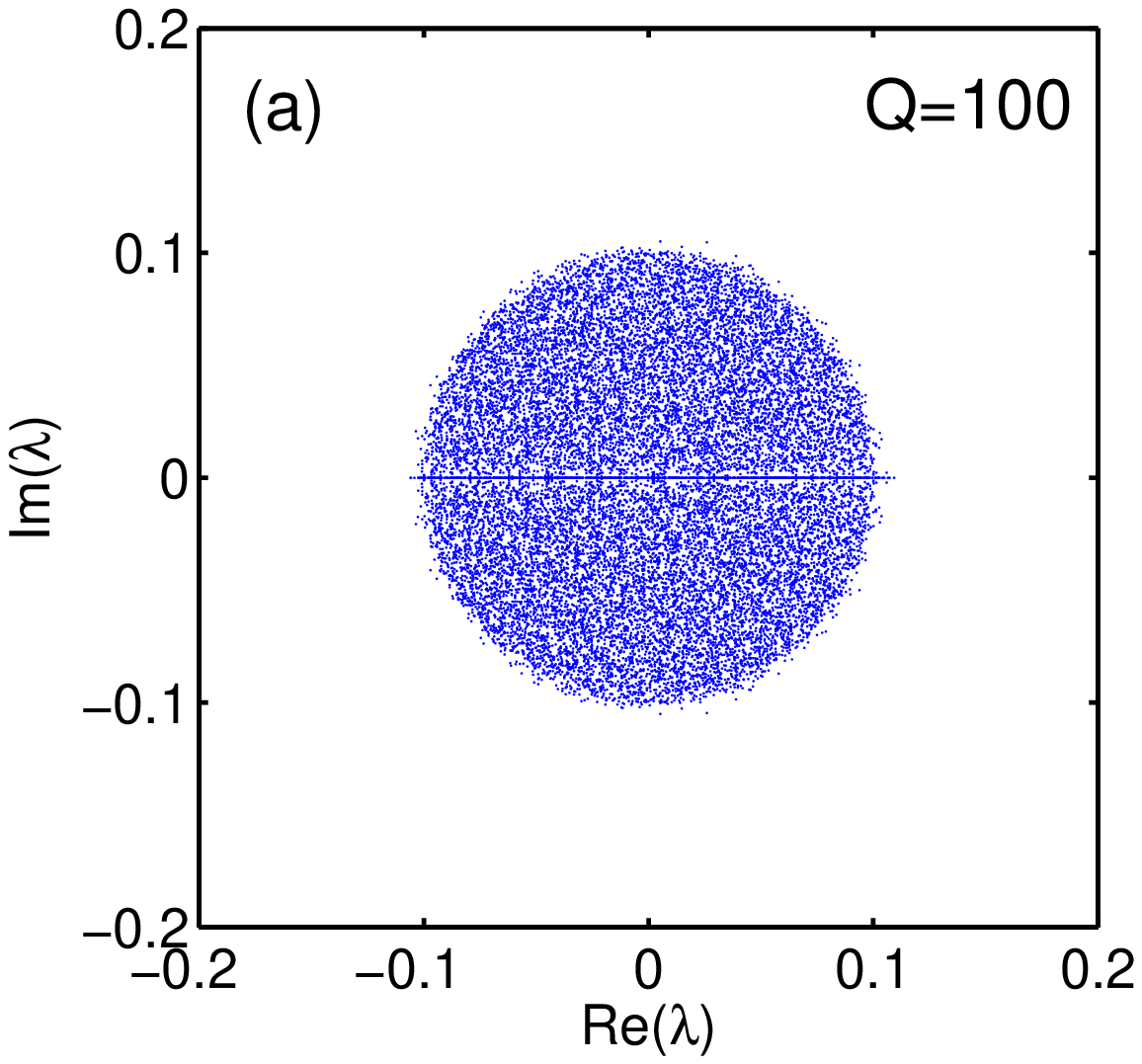}} & \resizebox{0.25\textwidth}{!}{\includegraphics{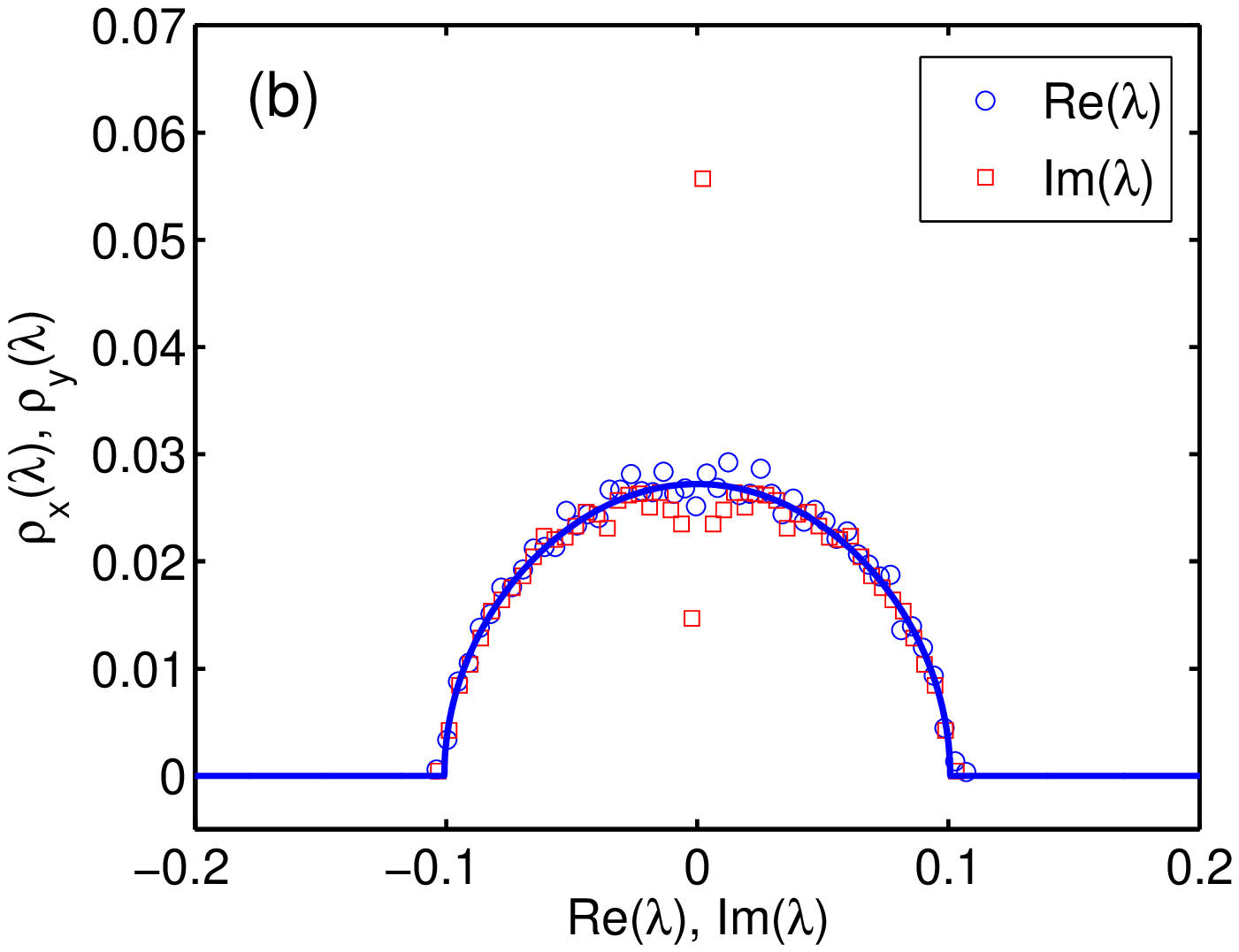} } \\
\resizebox{0.25\textwidth}{!}{\includegraphics{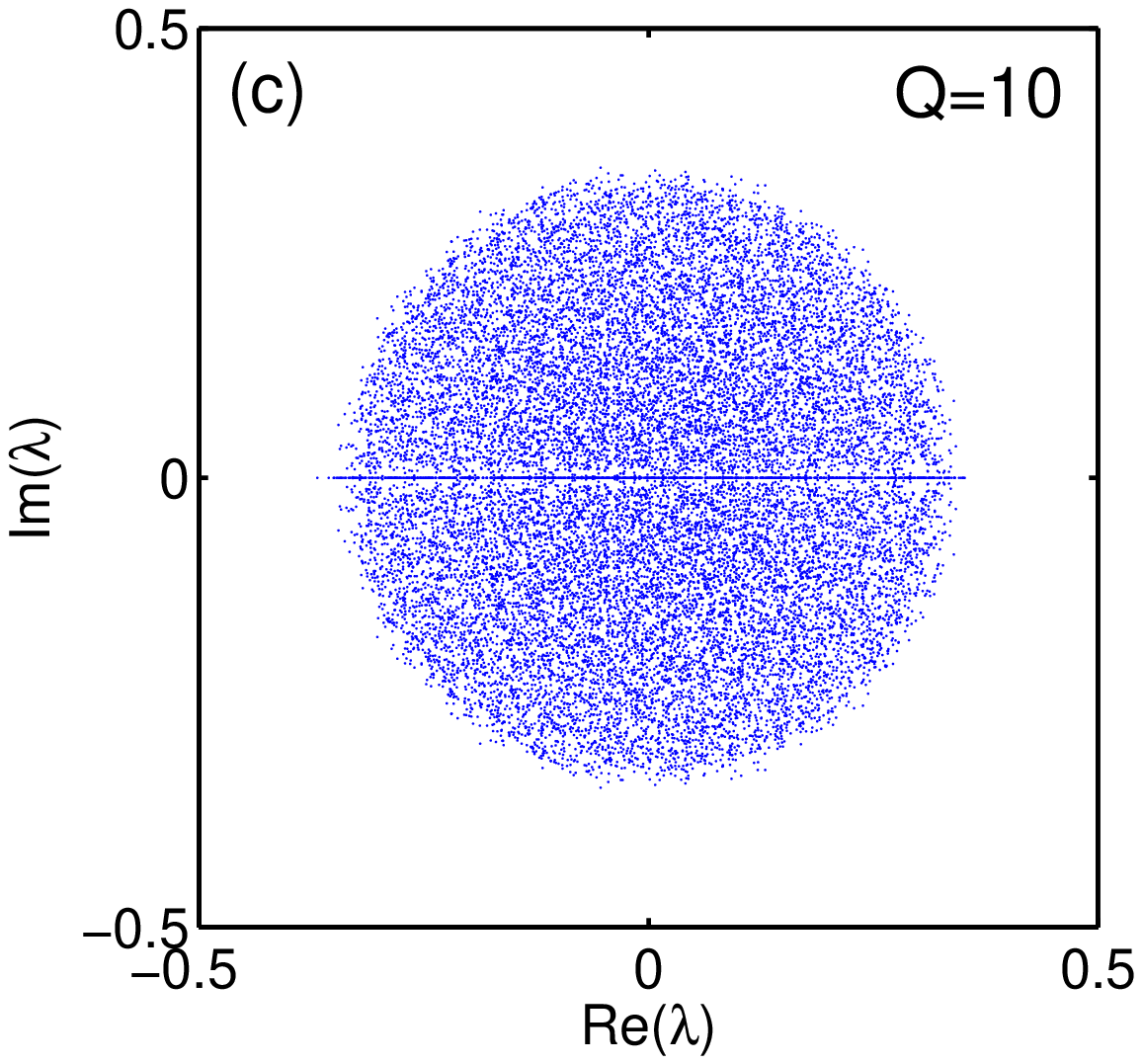}}  & \resizebox{0.25\textwidth}{!}{\includegraphics{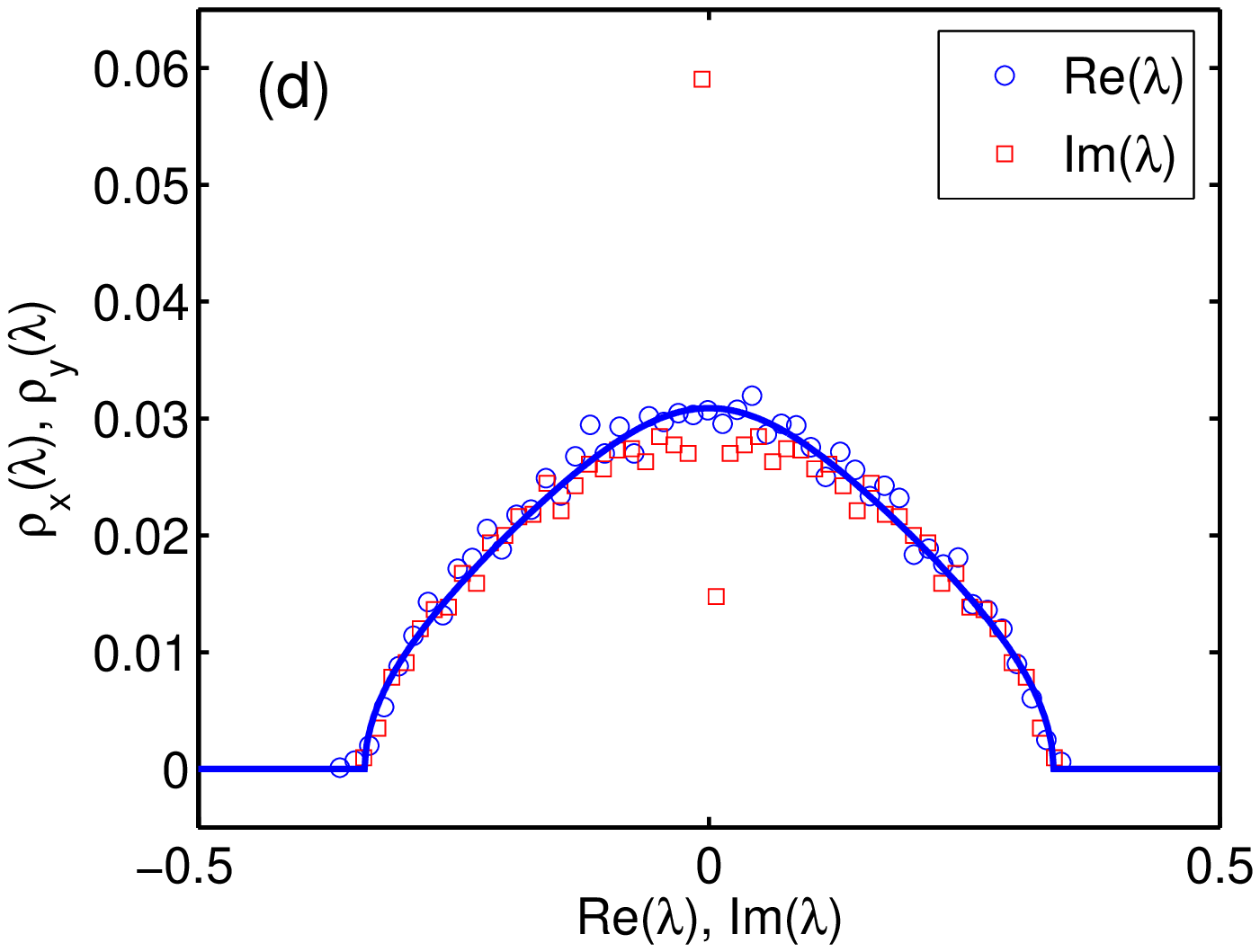} }  \\
\resizebox{0.25\textwidth}{!}{\includegraphics{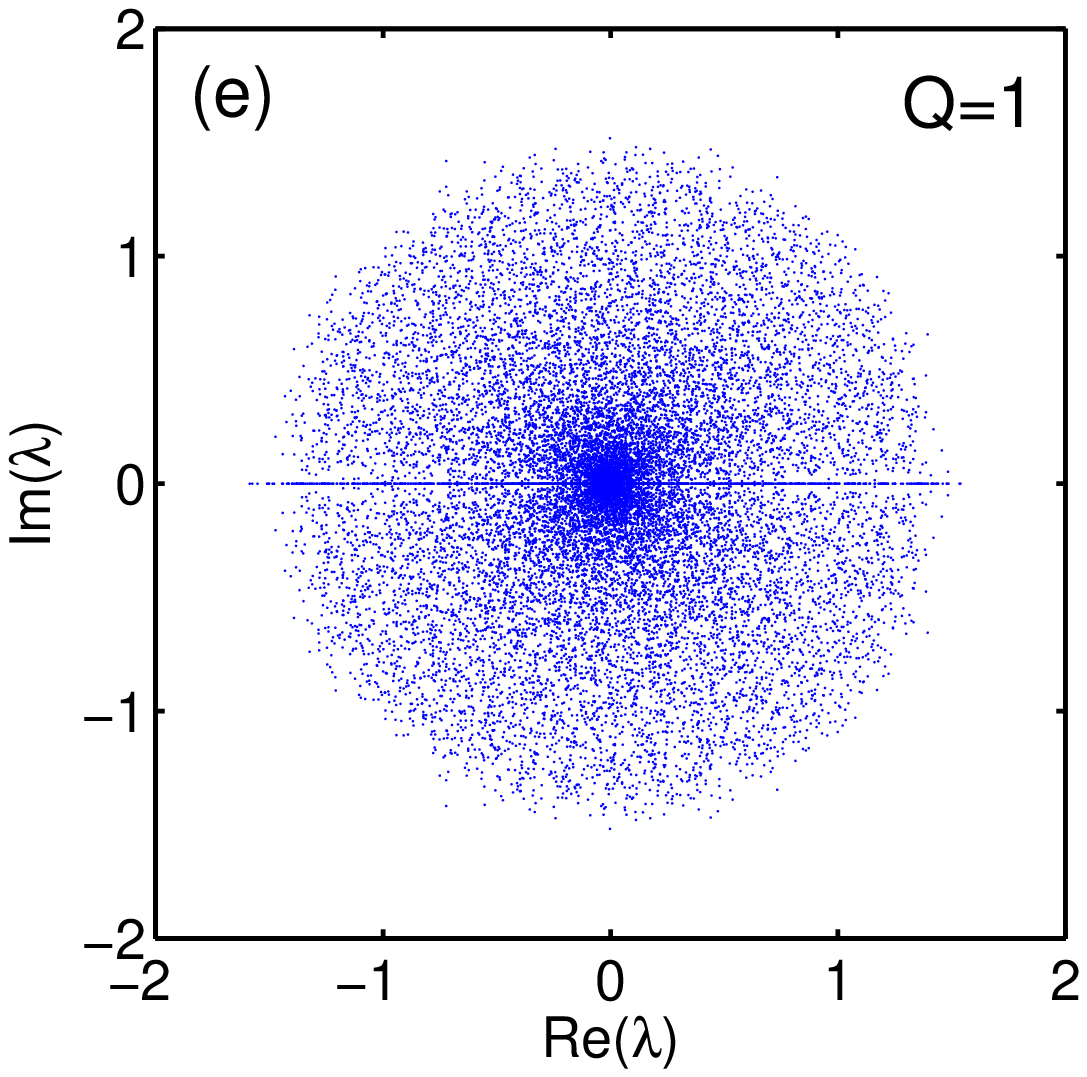}}   & \resizebox{0.25\textwidth}{!}{\includegraphics{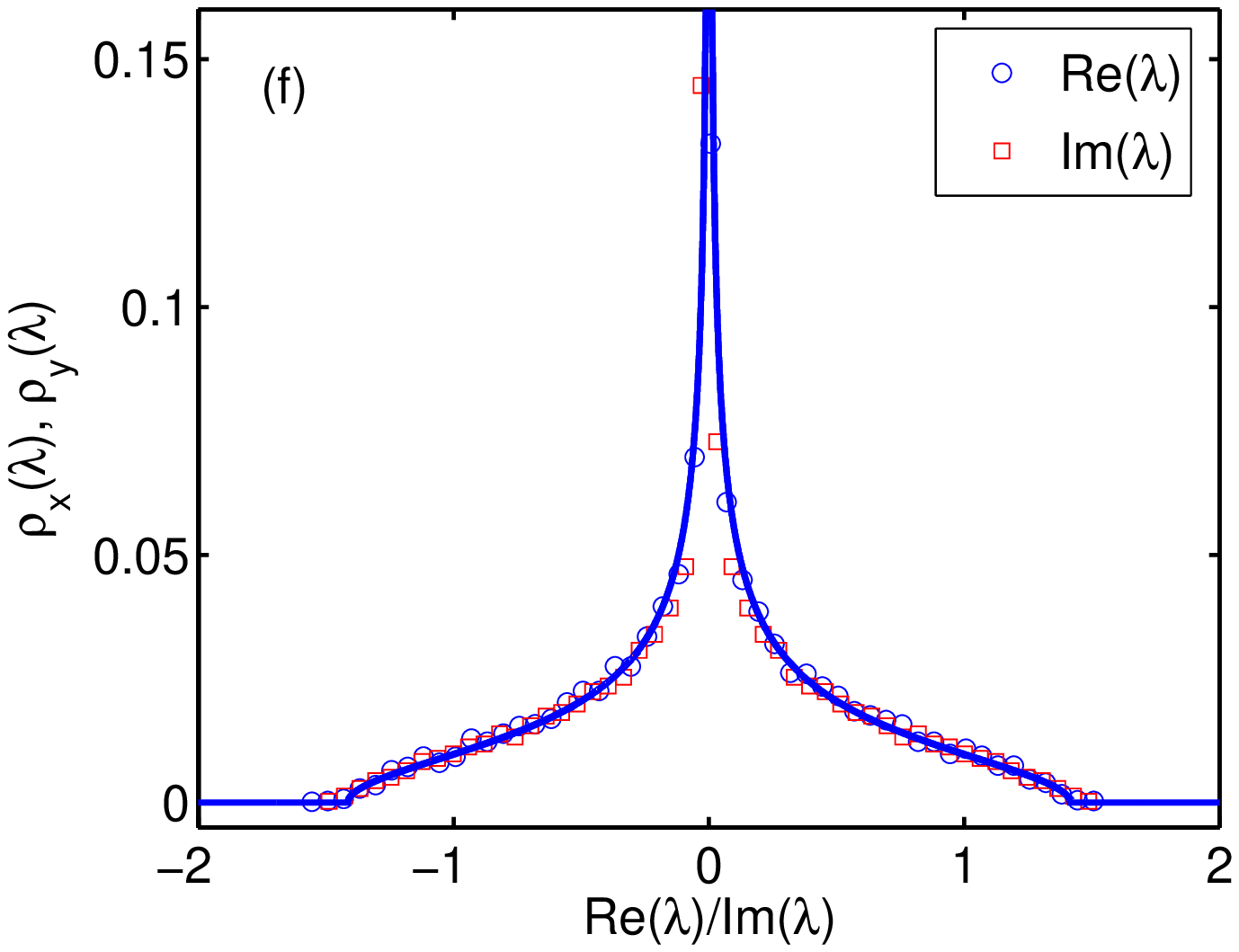}}    \\
\resizebox{0.25\textwidth}{!}{\includegraphics{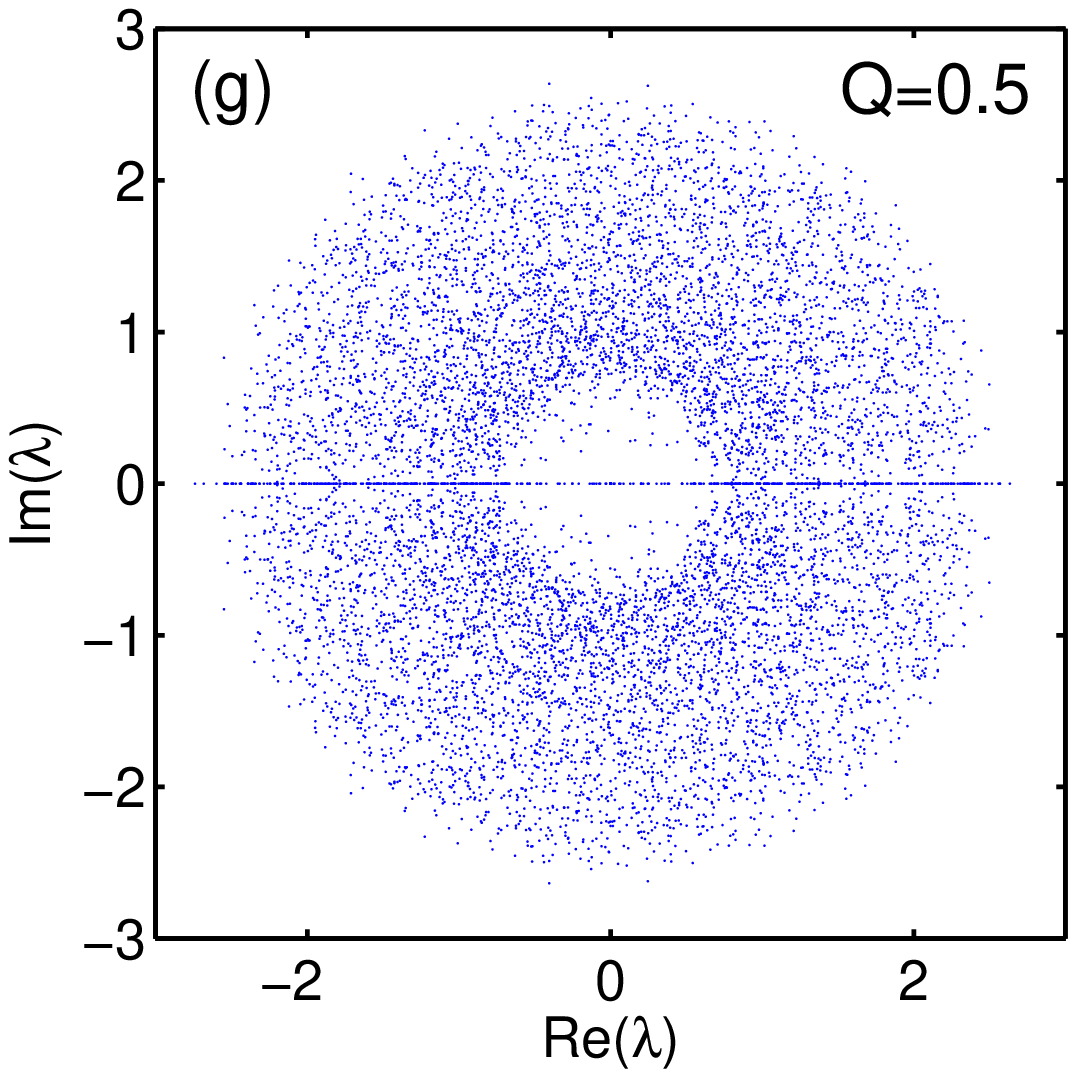}}   & \resizebox{0.25\textwidth}{!}{\includegraphics{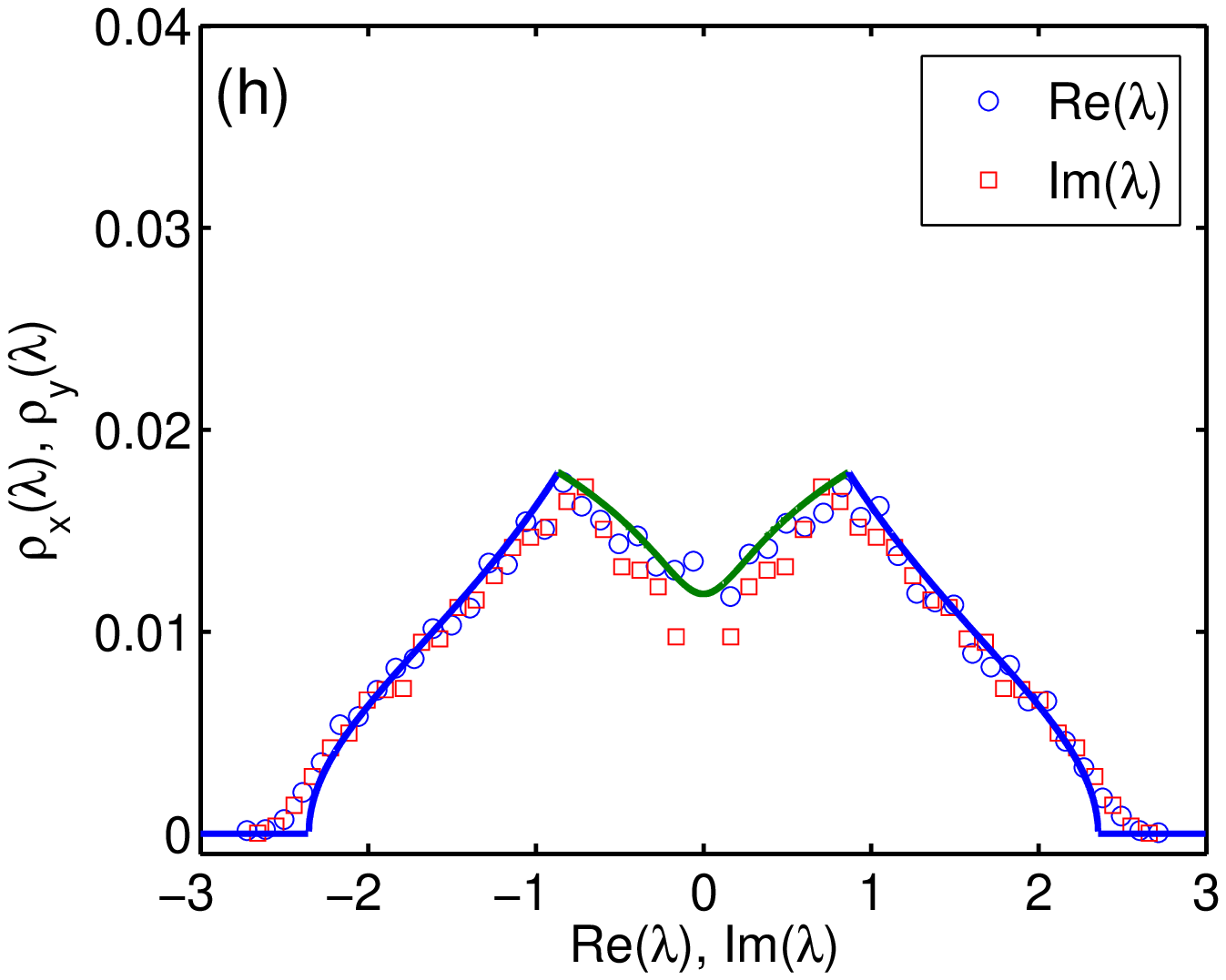}}    \\
\end{tabular}
\end{center}
\caption{ \label{pic0} 
Complex eigenvalue spectra of time-lagged  correlation matrices, obtained from 
random matrices ${\bf X}$. The entries of ${\bf X}$ are iid and Gaussian 
with unit variance.
In (a), (c), (e) and (g) the position of the eigenvalues is shown in the complex plane 
for  values of $Q\equiv \frac{T}{N}=100, 10, 1$ and $0.5$, respectively. 
The visibly enhanced density along the real axis is the finite-size effect mentioned in the text.
The right column shows the projections of the EVs onto the real and imaginary axis. 
The solid lines are the theoretically expected curves, 
which are numerical solutions to Eq. (\ref{hermitized}). 
Note in (h) that for this projection, the eigenvalue spectra
is composed of different solutions to Eq. (\ref{hermitized}) as $G(z)$ 
itself has a discontinuity. 
The divergence at $z=0$ is not shown for analytical curves associated with
$Q=100, 10$ and $0.5$.
}
\end{figure}
\end{center}

If there is no structural difference in the randomness of the symmetric and the
anti-symmetric part of matrix ${\bf C_{\tau}}$, the expression of 
Eq. (\ref{argument}) is equivalent under exchange of  $x$ and $y$ in the distribution sense, 
and Eq. (\ref{rho2}) will thus be a symmetric function in $x$ and $y$. 
Since we do not expect any direction in the complex plane 
being distinguished from any other in the limit $N\rightarrow\infty$, we 
conceive that the eigenvalue density resulting from (\ref{potential}) 
is a radial symmetric function, i.e., 
\begin{equation}
\rho(x,y)=\rho(r) \equiv \frac{1}{2 \pi r} \int_{\cal S} dz \rho(z) \, \delta(|z|-r)
\quad . 
\end{equation}
A more formal argument can be given via expanding the matrix $H_{ij}$ 
entering the potential $\phi$ \cite{privcomm}. Since the entries in $C_{ij}$ 
are typically smaller than one, 
$H_{ij}$ can be written as  $H_{ij}\approx{}|z|(A+\epsilon{}B)$. 
Here, $\epsilon$ is a small perturbation, $A=\delta_{ij}$ and 
$B=C^{ij}C^{ji}/|z|-\bar{x}(C^{ij}-C^{ji})+i\bar{y}(C^{ij}+C^{ji})$ with 
$\bar{x}=x/|z|$ and $\bar{y}=y/|z|$. We fix $|z|=1$ without loss of generality 
and write the determinant as a Taylor series,  
\begin{equation}
\label{expansion}
\begin{split}
\phi(x,y)&=-\frac{1}{N}\langle{}\ln\det(H_{ij})\rangle_c 
=-\frac{1}{N}\langle{}\mathrm{Tr}\ln(H_{ij})\rangle_c \\
&\approx{}-\frac{1}{N}\langle{}\mathrm{Tr}(B)-\mathrm{Tr}(\frac{B^2}{2})+\mathrm{Tr}(\frac{B^3}{3})- \cdots \rangle_c
\quad .
\end{split}
\end{equation}
Based on this series, we checked up to fourth order that this expansion indeed only 
leads to terms in $r$ for $N\rightarrow\infty$;   
we outline some aspects of the calculation in Appendix A.
We note that yet a different and probably even more powerful way of 
proving our conjecture would be to replace the determinant in Eq. (\ref{potential})
by Gaussian integrals and use the replica method to average over the 
distribution of the $C_{ij}$. 

If $\rho(r)$ is circular symmetric, the support $\cal S$ of the
eigenvalue-spectrum will be bounded by a circle and is thus definable via a
maximal
radius $r_{max}$. 
Since $r_{max}$
is governed by the
standard deviation of the underlying random matrix elements, one can compute the extent of the 
support of ${\bf C_{\tau}}$ by considering the support of symmetric ($r^S_{max}$) and anti-symmetric 
matrices ($r^A_{max}$). 
Let these be defined by  ${\bf C_{\tau}^S} \equiv \frac12 ({\bf C_{\tau}}+{\bf C_{\tau}}^T)$ and 
${\bf C_{\tau}^A} \equiv \frac{1}{2}({\bf C_{\tau}}-{\bf C_{\tau}}^T)$.
If we assume that the standard deviations of the symmetric and anti-symmetric matrices
are equal, $\sigma_S=\sigma_A$,  this 
implies that the standard deviation $\sigma$ of the 
matrix $C^{ij}_{\tau}$, will be $\sigma=\sqrt{2}\sigma_S/2$. Thus,
the support of ${\bf C_{\tau}}$ can be defined via a disc with radius
\begin{equation}
r^{\cal S}_{max}=\frac{1}{\sqrt{2}}r_{max}^S=\frac{1}{\sqrt{2}}r_{max}^A
\end{equation} 
The argument here is that the eigenvalue-density can be regarded as a 
log-gas \cite{forrester} which has only one degree of freedom for ${\bf
C_{\tau}^S}$ and ${\bf C_{\tau}^A}$, but two degrees of freedom for ${\bf
C_{\tau}}$, hence leading to $\sigma=\sqrt{2}\sigma_S/2$ instead of $\sqrt{2}\sigma_S$. 

Based on these relations and regarding the discussion of Eq. (\ref{argument}), 
it is sensible to conjecture that the projections 
of $\rho(r)$ onto the $x$-axis, denoted by $\rho_x(\lambda)$, 
and the projection onto the $y$-axis, $\rho_y(\lambda)$, 
are nothing but the rescaled spectra of the solution to the symmetric, 
$\rho^S(\lambda)$, and to the anti-symmetric problem, $\rho^A(y)$. 
To be more explicit, 
\begin{equation}
\label{projection}
\begin{split}
  \rho_x(\lambda) \equiv \rho(\mathrm{Re}(\lambda))=\int_{\mathcal{S}}\rho(r)dy  =  \rho^S(\sqrt{2}x)\\
  \rho_y(\lambda) \equiv \rho(\mathrm{Im}(\lambda))=\int_{\mathcal{S}}\rho(r)dx  =  \rho^A(\sqrt{2}y)
\end{split}
\quad , 
\end{equation}
where the integration extends over the support $\mathcal{S}$ in the complex plane. 
Although this conjecture might seem quite natural 
we shall provide numerical evidence for its correctness below. 

First, we note that the eigenvalue density of the symmetric problem can be obtained from the
well-known relation
\begin{equation}
\label{eigenvaluedensity}
  \rho^S(x)=\sum_{n}\delta(x-x_n)=
  \frac{1}{\pi}\lim_{\epsilon\rightarrow{}0}\left[\textrm{Im}(G^S(x-i\epsilon))\right] \quad.
\end{equation}
For a radial symmetric problem, of course,  $\rho^S \sim{} \rho^A$.
The main idea of this work is now to note that one can use the following technique
to actually determine the radial symmetric density $\rho(r)$:

Since the rescaled eigenvalue density of the symmetrized problem $\rho^S({\sqrt{2}x})$ is
nothing but the projection of $\rho(r)$ onto the real axis,
Eq. (\ref{projection}), it can  be written as the
Abel-transform \cite{bracewell}, 
\begin{equation}
\rho^S(\sqrt{2}x)=2\int_{{x}}^\infty{}\frac{\rho(r)r}{\sqrt{r^2-{x}^2}}dr \quad , 
\end{equation}
of the  radial density $\rho(r)$.
One can then reconstruct the desired eigenvalue spectrum \emph{exactly} (in the limit $N\rightarrow\infty$)
via the inverse Abel-transform, and thus via the cuts of the Greens function
of the symmetric problem, 
\begin{equation}
\label{abelsolution}
  \rho(r)=-\frac{1}{\pi}\int_r^\infty{}
  \frac{\frac{d}{dx}\lim_{\epsilon\rightarrow{}0}\left[\textrm{Im}(G^S_\tau(\sqrt{2}x-i\epsilon))\right]}
  {\sqrt{x^2-r^2}}\mathrm{d}x \quad. 
\end{equation}
Here, we have made use of Eq. (\ref{eigenvaluedensity}). 
Since Eq. (\ref{abelsolution}) can be problematic
if evaluated numerically, 
we also specify 
a form which exploits the Fourier-Hankel-Abel cycle \cite{bracewell}
\begin{equation}
\rho(r)=2\pi\int_0^{\infty}q\mathrm{J}_0(2\pi{}rq)
\int_{-\infty}^{\infty}
\rho^S(x)
e^{-2\pi{}ixq}\mathrm{d}x\,
\mathrm{d}q \quad ,
\end{equation}
where $\mathrm{J}_0(x)$ denotes the zeroth-order Besselfunction. We also note,
that yet another method of determining $\rho(r)$ is the evaluation of
the inverse Radon-transform of $\rho^S(\sqrt{2}x)$.

Equation (\ref{abelsolution}) applies for \emph{any} radial symmetric eigenvalue
density in the limit $N\rightarrow\infty$ and allows for a calculation of the
eigenvalue density in the complex plane via a method of exact reconstruction based on
the eigenvalue density of the symmetrized (or anti-symmetrized) problem. 
Typically, the solution of the symmetric problem will be valid only in the $N\rightarrow\infty$ limit. 
Thus, although the Abel-inversion gives an exact result, discrepancies may occur because of finite-size effects.
Before turning to the specific problem of lagged correlation matrices we refer to Appendix B, where 
-- as a specific and prominent example -- we show the almost trivial case of deriving the 
density of real asymmetric random 
matrices  (without 'imposed structure') 
\cite{crisantisommers} directly from Wigner's celebrated semicircle law.

\begin{figure}
\begin{center}
 \resizebox{0.4\textwidth}{!}{\includegraphics{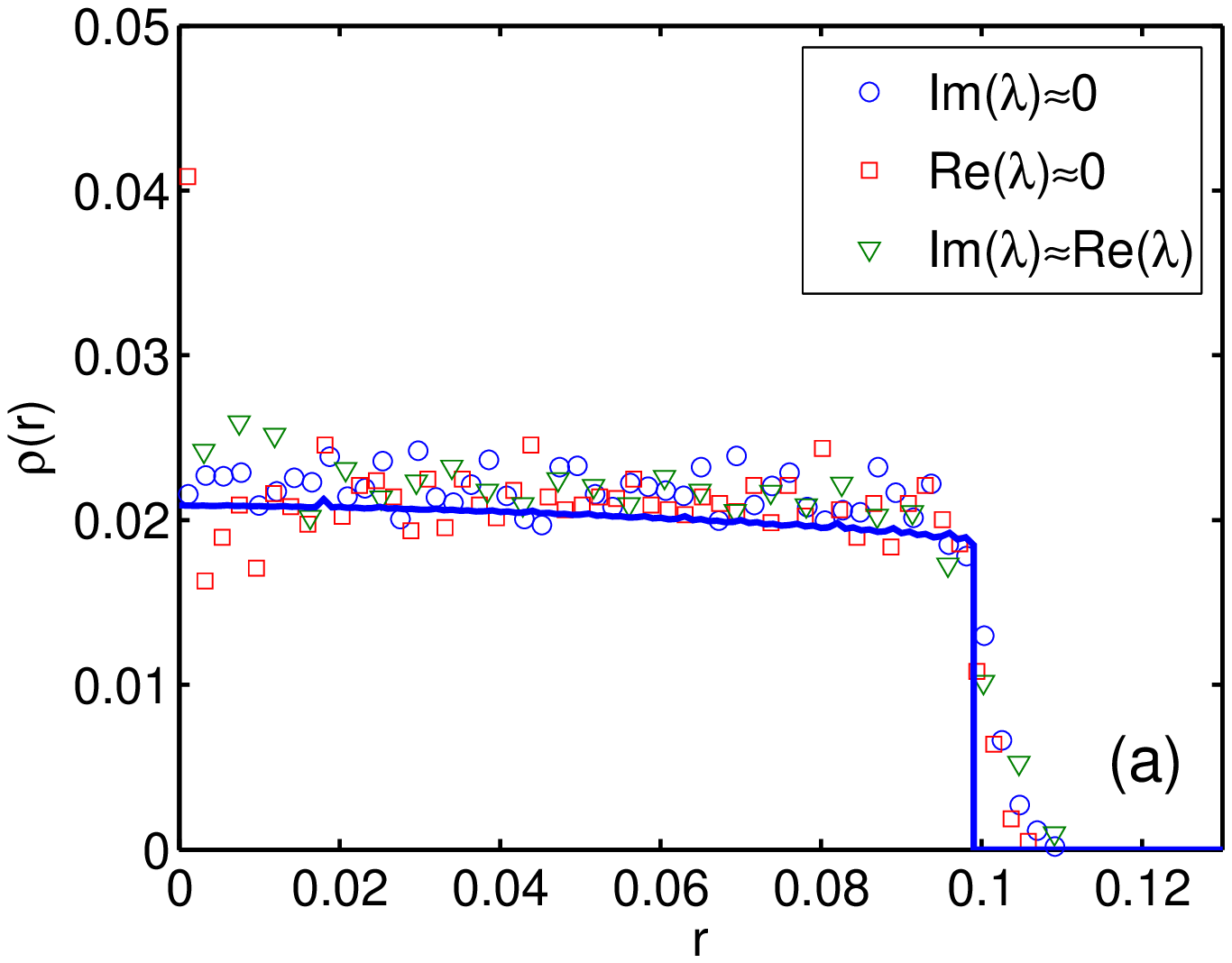}}
 \resizebox{0.4\textwidth}{!}{\includegraphics{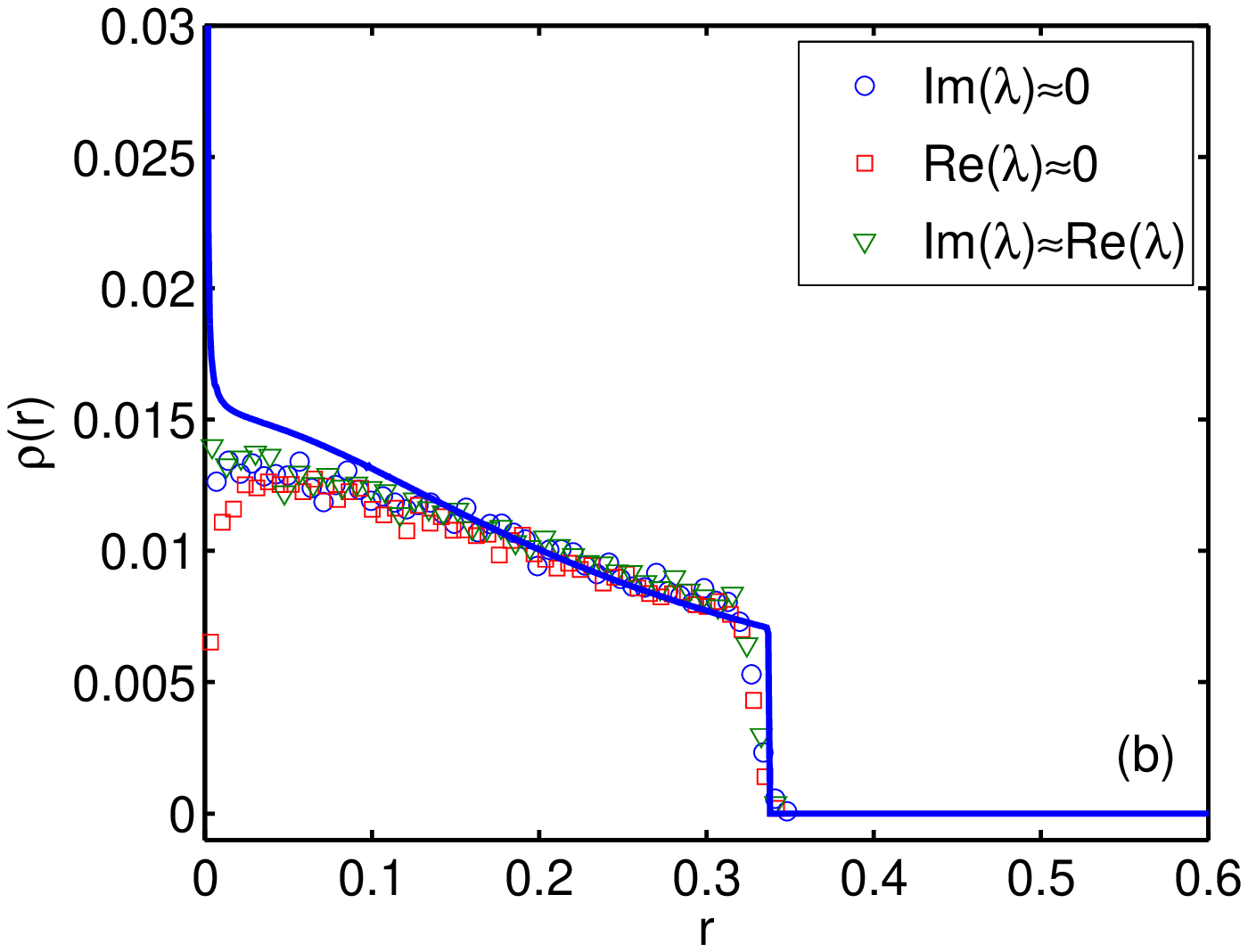}}
 \resizebox{0.4\textwidth}{!}{\includegraphics{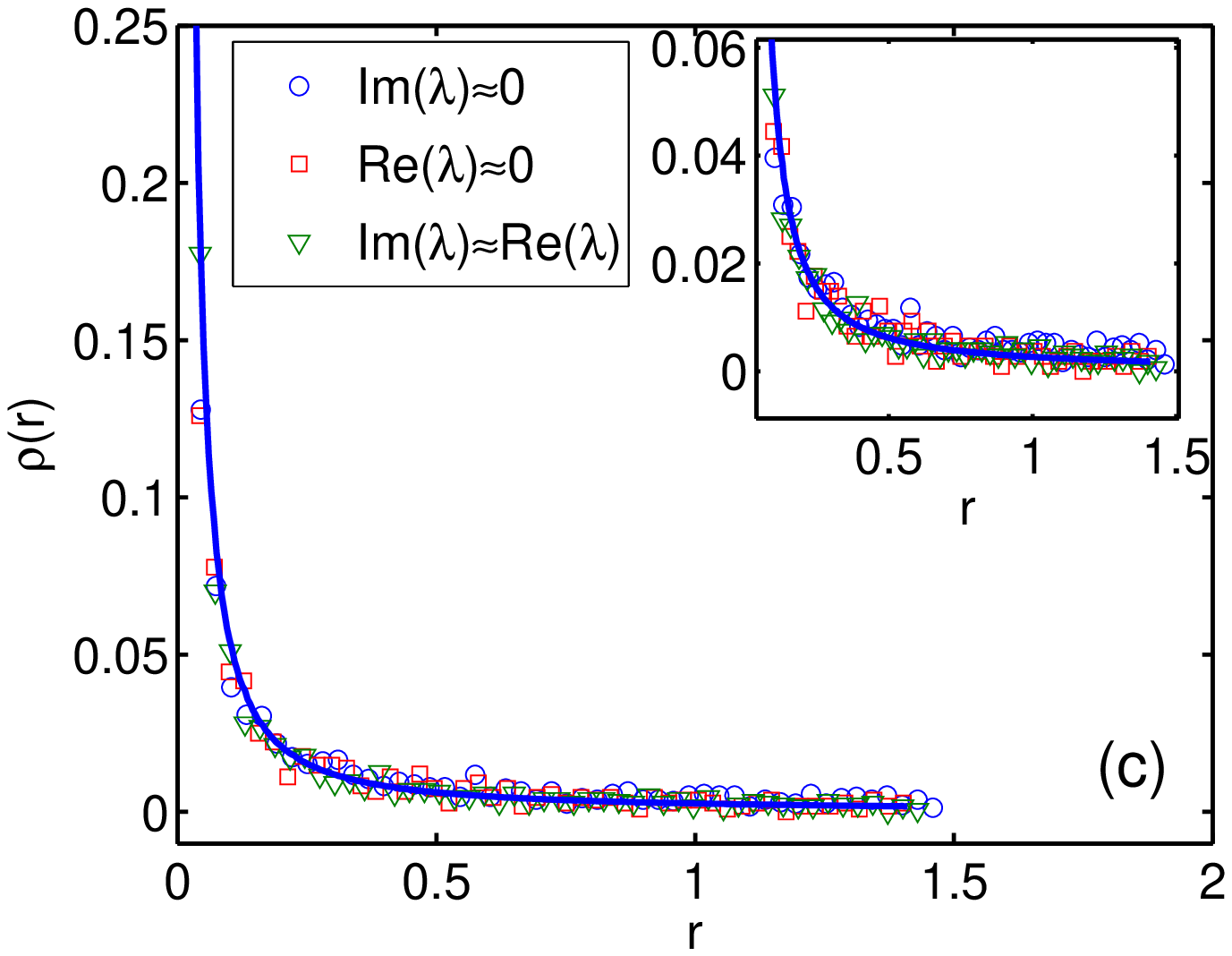}}
\end{center}
\caption{Radial eigenvalue densities approximated via simulations along different
directions (real axis, imaginary axis and the diagonal in the complex plane.
Numerical data for finite matrices is compared with the solution of the inverse Abel-transform.
(a) Q=100 
(b) Q=10
(c) Q=1; the inset shows a detail of the curve.}
\label{abelcuts} 
\end{figure}

\subsection{Application to lagged correlation matrices}

We now turn to our specific problem of determining the eigenvalue density of ${\bf
C}_{\tau}$. What is left is to confirm the validity of our conjecture, 
Eq. (\ref{projection}), and to show, that -- as a consequence -- Eq. (\ref{abelsolution}) 
gives an approximation to the radial eigenvalue distribution, $\rho(r)$. 
To start, we can refer to existing literature on the symmetric problem:
It has been shown \cite{delay_corr,burda_financial}, that the Greens function, 
$G(z)$ of the symmetric problem ${\bf C_{\tau}}^{S}=\frac{1}{2T}
{\bf X}({\bf D}_{\tau}+{\bf D}_{-\tau}) {\bf X}^T$ is given by
\begin{equation}
\label{hermitized}
\begin{array}{c}
\frac{1}{Q^3}z^2G^4(z)-2\frac{1}{Q^2}(\frac{1}{Q}-1)zG^3(z)-\\
\frac{1}{Q}(z^2-(\frac{1}{Q}-1)^2)G^2(z)\\
 +2(\frac{1}{Q}-1)zG(z)+2-\frac{1}{Q}=0
\quad , 
\end{array}
\end{equation}
with $Q \equiv T/N$ playing the role of a information-to-noise ratio. Note,
that this equation is independent of a specific value for $\tau$ and is valid
for any value of it \cite{burda_financial}.

We note, that -- in a calculation analogous to the
one in \cite{burda_financial} -- it is easy to show that 
the Greens function pertaining to the asymmetric problem
follows exactly the same equation, 
which reaffirms circular symmetry.
Based on Eq. (\ref{hermitized}) one can calculate  $\rho_x(\lambda)$ 
by using Eqs. (\ref{eigenvaluedensity}) and (\ref{projection}).

Figure \ref{pic0} shows (simulated) spectra of ${\bf C_{\tau=1}}$ as defined by Eq. (\ref{Cdefinition}) with
iid entries in the columns of ${\bf X}$, for various values of $Q$. Note, that for $Q<1$ the shape of the boundary
of eigenvalues in the complex plane
changes from a disk to an annulus (see e.g. \cite{zee} for a discussion of 
disc-annulus phase transition in the case
of \emph{non-hermitian matrix} models).
We immediately recognize that eigenvalues are enhanced along the real axis and that, 
as a consequence, the density is lower in the 
vicinity of the real axis. This can be attributed to a well-known finite-size effect, 
already discussed in \cite{crisantisommers,crisantisommers2}. Of course, this effect 
implies that circular symmetry is not fully fulfilled for finite matrices of the GinOE. 
Thus, we also expect to observe some discrepancies between the theoretical 
results based on the Abel-transform and the empirical densities of finite, 
lagged correlation matrices based on random data. 

In our concrete case, the prediction of the projections $\rho_x$ and $\rho_y$ (blue lines, obtained from 
Eq. (\ref{eigenvaluedensity}) and Eq. (\ref{hermitized}))
depicted in the right column of Figure \ref{pic0} is in good 
agreement with the numerical data for the real parts of the eigenvalues ($\rho_x$). For the projection of the complex parts ($\rho_y$)
we recognize that there is a slight deviation from the prediction (due to the enhanced density along the real axis).
We also checked projections with data obtained via rotating all the
individual eigenvalues in the complex plane for different angles. Apart from some minor effects attributable to 
the inhomogenity around the real axis we found no
significant discrepancies. 
We also note that the simulated data did not show any significant discrepancies
when taking different values of $\tau$ which is again in agreement with the theoretical anticipation.

Turning towards the point of reconstructing the radial eigenvalue density, 
the function to be transformed ($\rho^S(\sqrt{2}x)$ or $\rho^A(\sqrt{2}y)$) 
may be  evaluated exactly (with some effort) for the symmetric case from 
Eq. (\ref{eigenvaluedensity}) and Eq. (\ref{hermitized}). The remaining integral
Eq. (\ref{abelsolution}) will, however, be hard to solve in general. 
Nonetheless, we are able to solve the case $Q=1$ analytically and obtain 
the exact formula for the eigenvalue density, 
\begin{equation}
\begin{split}
\rho_{Q=1}(r)=
&\frac{1}{K}
\left[
2^{3/4}{}3{}r\Gamma\left(\frac{5}{4}\right)\Gamma\left(\frac{5}{4}\right)\Phi_2^1\left(\frac{1}{4},\frac{5}{4},\frac{3}{2},\frac{\lambda^2}{2}\right)
\right.\\
&\left.
-2^{1/4}\Gamma\left(-\frac{1}{4}\right)\Gamma\left(\frac{7}{4}\right)\Phi_2^1\left(-\frac{1}{4},\frac{3}{4},\frac{1}{2},\frac{\lambda^2}{2}\right)
\right]
\quad , 
\end{split}
\end{equation}
with $K \equiv 6\sqrt{\pi^5{}r^3}$. 
Here, $\Gamma(x)$ denotes the Gamma function and $\Phi_2^1(a,b,c,z)$ the hypergeometric
function; the derivation is briefly
summarized in Appendix C.
Note that $\lim_{Q\rightarrow{}0}G^S_{Q}(z)=\frac{1}{z}$, whereas
for $Q\rightarrow\infty$ we
expect the Greens function and the eigenvalue density to converge to those of a
random real asymmetric matrix without specific structure, i.e. a flat
eigenvalue-density in the sense of 
\cite{crisantisommers}.

We were not able to derive closed expressions for other values of $Q$, since already the solution of Eq. (\ref{hermitized})
results in lengthy expressions.
In these cases we computed the integral Eq. (\ref{abelsolution}) numerically. 
The results are depicted in Figure \ref{abelcuts} for $Q=100$, $Q=10$ and $Q=1$.
The theoretical predictions are accompanied by data obtained from performing cuts along various directions
of the spectra $\rho(x,y)$ from Fig. \ref{pic0}, namely along the x-axis, the y-axis and along the
diagonal direction, i.e. $\mathrm{Re}(\lambda)=\mathrm{Im}(\lambda)$.
We performed these cuts numerically via calculating the density in narrow strips along the different directions.
The theoretical prediction catches the different experimental densities 
very well. Especially for $Q=100$ and $Q=1$ results are consistent with the predictions to a high degree. 
For $Q=10$ we observe some discrepancies for values $r<0.1$. 
We think that these are very probably associated with the finite-size effect of enhanced eigenvalue density along the real
axis discussed above. Actually, a closer investigation of this effect and a comparison with the solution found in \cite{edelman} would be interesting to do but remains outside the scope of the present work. 


\section{Empirical Analysis}
\label{sec3}

With a theoretical concept of, and some specific knowledge about, the eigenvalue-spectra 
of time-lagged correlation matrices, we now turn to 
actual financial data and study empirical lagged correlation matrices ${\bf C_{\tau}}$.

\subsection{Data}
We analyze 5 min data of the S\&{}P500 in the time period of Jan 2 2002 -- Apr 20 2004.
The time-series were cleaned, corrected for splits and synchronized. 
In particular, days where trading took only place in 'limited' form ('half-days' etc.) 
have been removed (this includes the dates Sep 11 2002, Dec 26 2003, Jan 19 2004, Feb 16 2004).
Additionally, all assets in which more that $1.5\%$ of data were missing and/or assets which were not
quoted over the full time-frame have been removed.
After cleaning, the  data set ${\bf X}$  consisted of $N=400$ time-series at $T=44720$ observation
times each. The empirical time-series and its distribution-functions showed the usual 
'stylized facts' of high-frequency stock-returns (fat-tails, clustered volatility, etc.).
Of course, also the well-known structure of correlation matrix element distribution at equal times 
was found to be present in the data (not shown). 
For the remainder of the paper, we fix $\tau=1$, i.e. a five minute shift, and $T=44720$, 
if not stated otherwise.
From ${\bf X}$ we construct two surrogate data sets, one by removing the market mode, the other by 
a scrambling of data.
As $\tau=1$ remains unchanged during the rest of the paper, we will 
occasionally drop the subscript, ${\bf C}_1={\bf C}$.

\subsubsection{Market mode removed data}
It is well known that the spectrum of equal-time correlations is 
dominated by a single very large eigenvalue which can be attributed to the so-called 'market-mode', 
see e.g. \cite{oldlaces,plerou, bouchaud_book}.
Removing the 'market mode' is thus approximately equivalent to removing the movement of the 
'index' of a given universe from the individual assets. 
We define the market return (the index) by $r^{m}_t= \sum_{j=1}^{N} v_{1j} r^j_t$, 
where $v_{1j}$ is the eigenvector associated with the largest eigenvalue
$\tilde{\lambda}_1$ of the empirical covariance matrix at \emph{equal times}, i.e. $\tau=0$.
To remove this market mode from the data we simply regress in the spirit of the CAPM 
\begin{equation}
\label{model}
r^i_t=\alpha^i+\beta^i r^{m}_t+\epsilon^i_t \quad , 
\end{equation}
where the residuals $\epsilon^i_t$ carry what is left of the structural information 
in the data; we denote this data set by  ${\bf X^{res}}$, its elements being
$X^{\rm res}_{it}=\epsilon_t^i$.

\subsubsection{Scrambled data}

A scrambled version ${\bf X^{scr}}$ is generated by a random permutation of all elements of ${\bf X}$. 
This destroys all correlation structure but has exactly the same distributions as the original data.
Correlation matrices from ${\bf X^{scr}}$ should -- up to potential non-Gaussian effects in 
the distributions -- correspond to the
developments in Section \ref{sec2}. We checked that the support of the
eigenvalue-spectra pertaining to the lagged correlation matrices  -- which
will be the quantity used for identifying deviating eigenvalues -- indeed
resembles the value $r_{max}$ of the Gaussian case discussed in Section
\ref{sec2}.
A treatment of the exact spectra of lagged correlation matrices of random Levy distributed data
(see e.g. \cite{burda_financial,Bochaud_neu,Burda_neu} for the case of
equal-time covariance matrices) is beyond the scope of the present work.

\begin{center}
\begin{figure}
\begin{center}
\resizebox{0.4\textwidth}{!}{\includegraphics{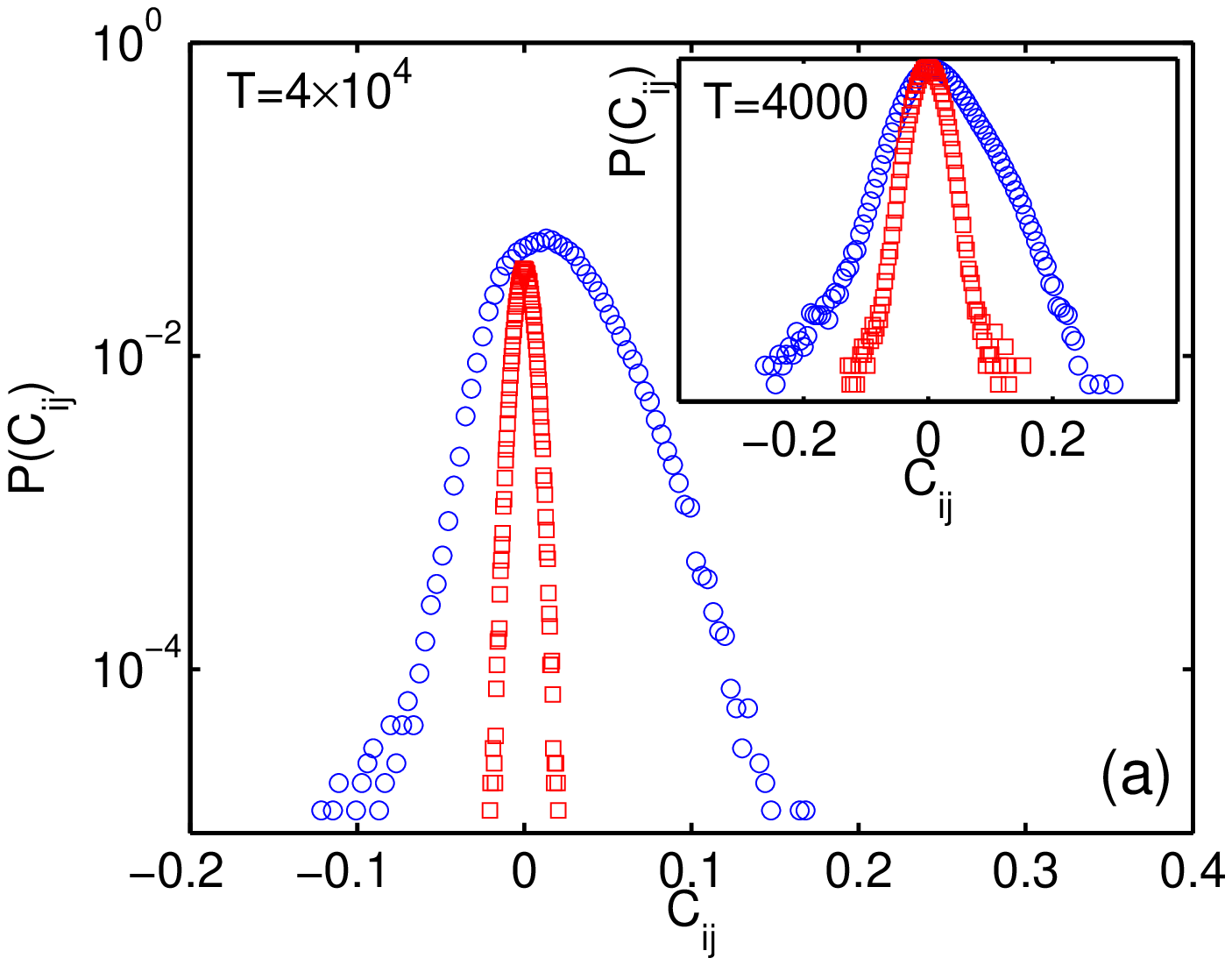}}
\resizebox{0.4\textwidth}{!}{\includegraphics{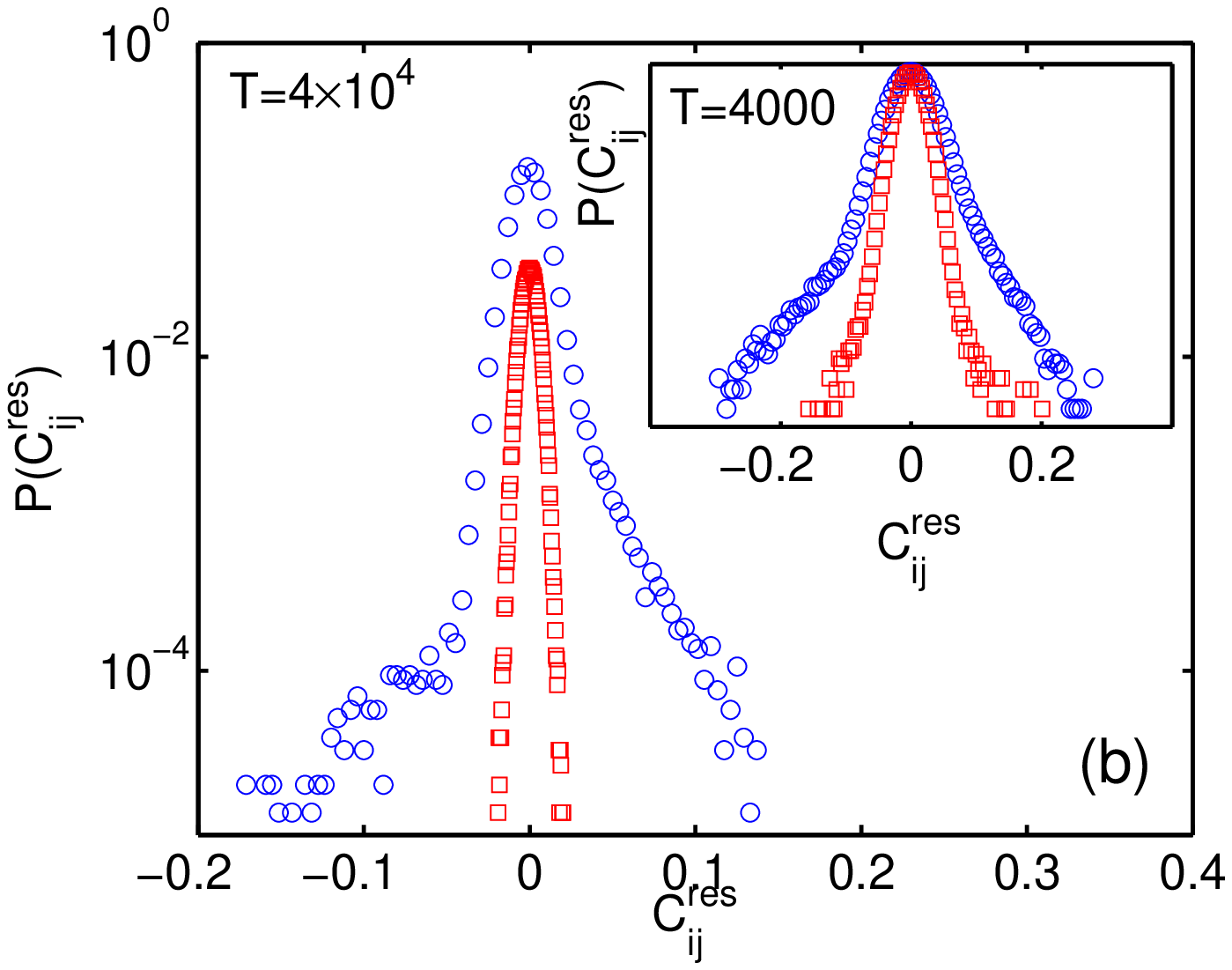}}
\end{center}
\caption{
(a) Empirical distribution $P(C_{ij})$ of the lagged
correlation matrix elements $C_{ij}$ for a sampling period of 
$T=40000$ and $T=4000$ (inset). Circles represent empirical data,  
red squares the situation for scrambled data from ${\bf X^{scr}}$. 
(b) shows the same for the  removed market case,  
i.e. from ${\bf X}^{res}$. Individual frequencies are normalized by the summed
frequencies for each plot.
}
\label{pic22}
\end{figure}
\end{center}

\subsection{Empirical time-lagged financial random matrices}

In Fig. \ref{pic22} we show the distribution of matrix elements  $P(C_1^{ij})$  (circles) 
of the empirical correlation matrix ${\bf C_1}$, based on ${\bf X}$ 
(a), and ${\bf X^{res}}$ (b). Squares show the results for the scrambled data 
${\bf C_1^{scr}}$. The inset shows the result for a shorter sampling time of $T=4000$. 
Clearly, there is 'significant' correlation in the data in both cases, contrasting
the Gaussian prediction of the efficient market hypothesis. 
The effect of varying the time-difference aspect of lagged correlations
has been carefully studied in \cite{kertesz2}, and  we shall not discuss this
issue here. However we point out, that -- as expected -- the lagged correlations at $\tau=1$ were larger than
for values of $\tau>1$, which is fully conforming with the findings of \cite{kertesz2}.
We also mention that we see that correlations typically decrease with decreasing 
observation frequency  (comparing 5 min data with hourly returns), 
but still remain well above the scrambled case (not shown).

The situation for the market removed data ${\bf X^{res}}$, (Fig. \ref{pic22} (b), shows 
that lagged correlations are not distributed according to the 
efficient market hypothesis as well. 
The frequency of higher values of $C_1^{ij}$ is slightly reduced and the curve
has significantly changed shape. In the semilogarithmic plot of Fig. \ref{pic22}, 
the positive regime is clearly not following a square-polynomial curvature,
but rather an exponential one. This also applies to the data sampled from
$T=4000$ subperiods, depicted in the inset of Fig. \ref{pic22}. Both empirical
distribution functions also exhibit clear non-random negative autocorrelations which are the predominant 
source of the non-Gaussian tails for negative entries.

\begin{center}
\begin{figure}
\begin{center}
\resizebox{0.5\textwidth}{!}{\includegraphics{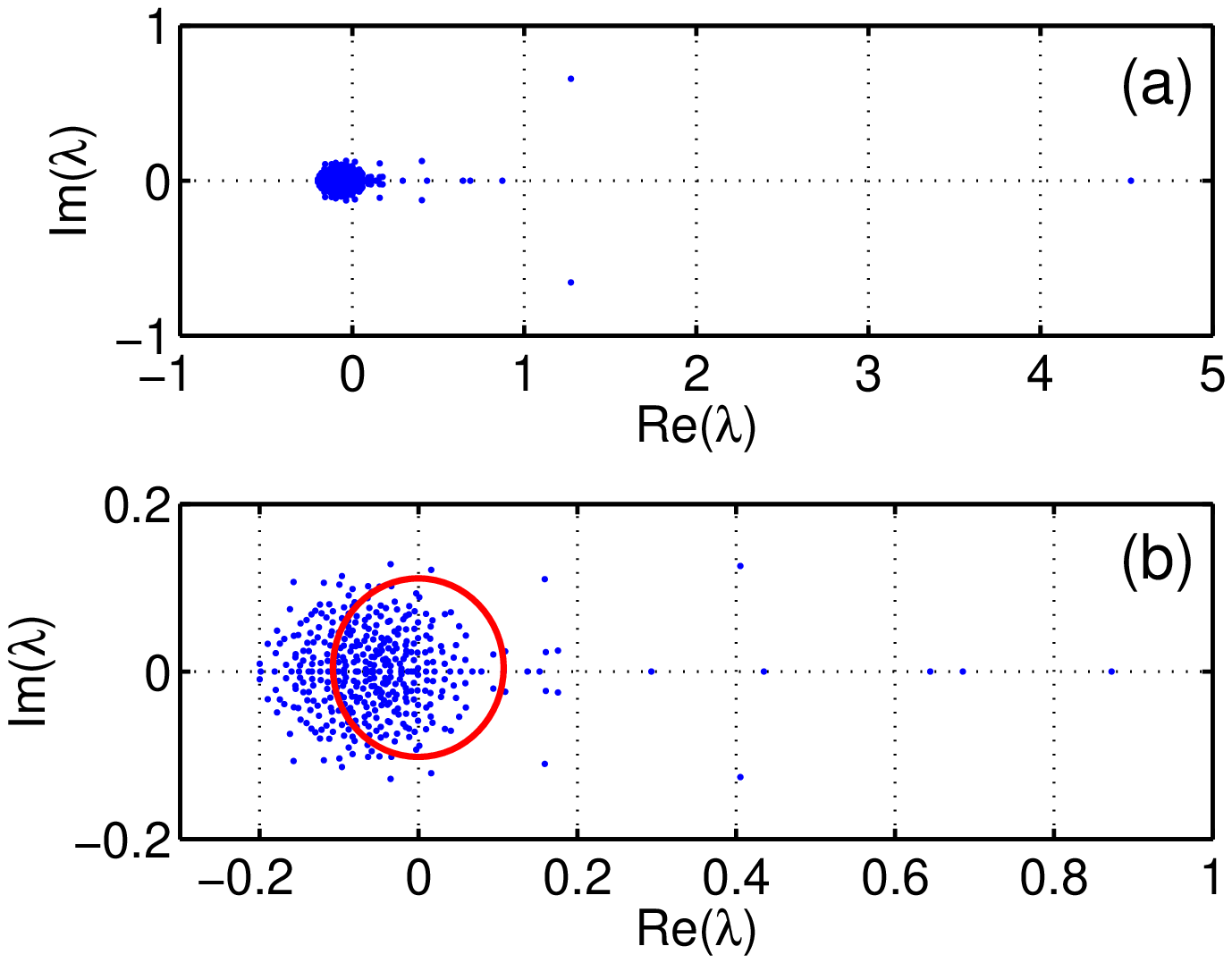}}
\resizebox{0.5\textwidth}{!}{\includegraphics{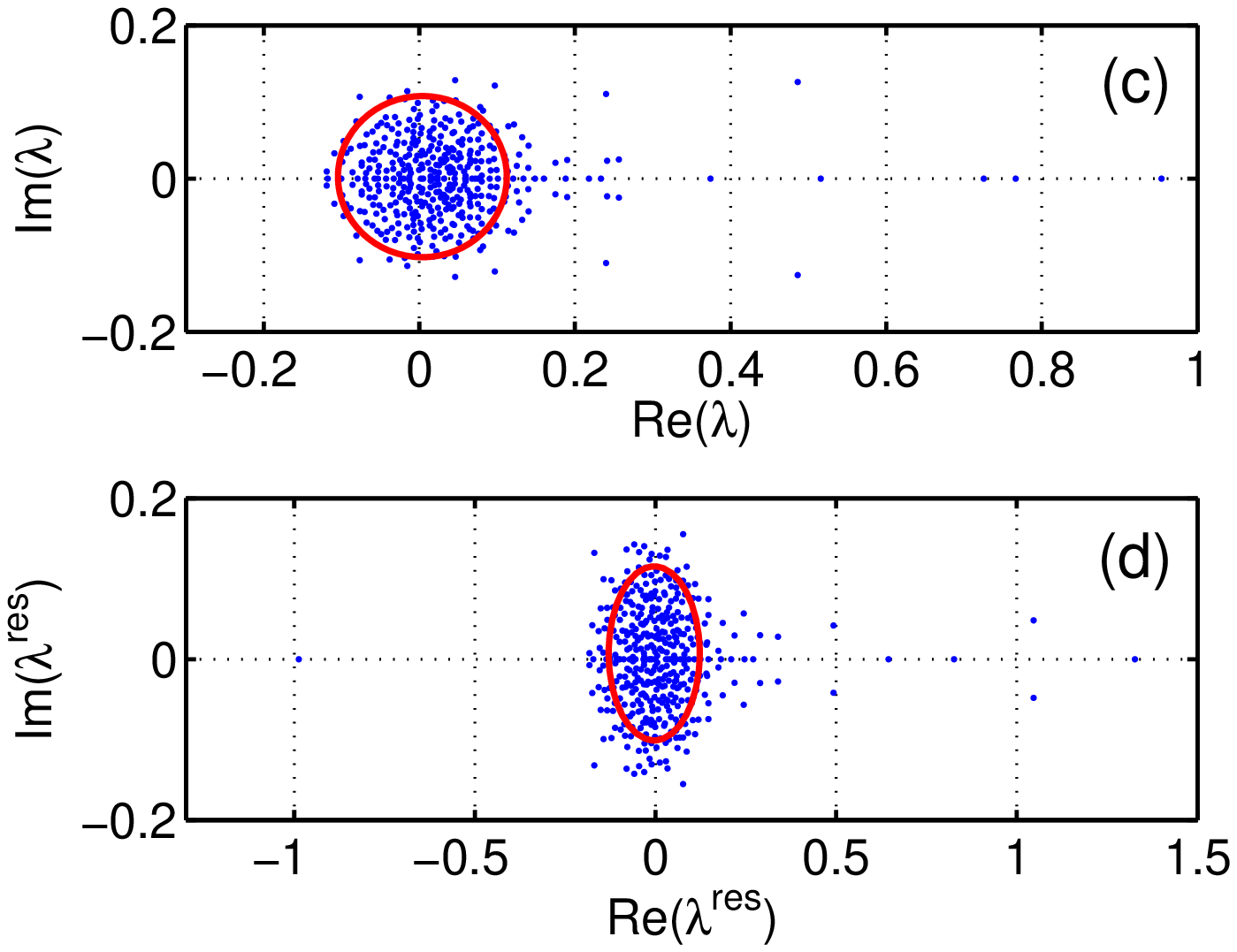}}
\end{center}
\caption{Eigenvalue spectra of lagged correlation matrices from 5 min 
S\&P500 data. 
(a) shows the full spectrum with one  very large deviation  on the real axis ($\lambda_1 \sim 4.6$),  
and a large departing eigenvalue pair $\lambda_2=\lambda_3^*$. 
(b) is a detail, clearly showing that the spectrum is shifted with respect to the 'bulk-disc'. 
(c) spectrum corrected for displacement $d$ as discussed in the text. 
(d) is the eigenvalue spectrum based on the market removed data, ${\bf X^{res}}$,  also after 
displacement correction.
The circles in plots (b)--(d) indicate the theoretical support discussed in Section \ref{sec2}.
}
\label{nonregressed2} 
\end{figure}
\end{center}

\subsubsection{Eigenvalue spectra}

We now proceed to the analysis of empirical eigenvalue spectra of the
financial data.
Figure \ref{nonregressed2} (a)-(c) shows the eigenvalue spectrum obtained from
${\bf {C}}$ at various stages. In Fig. \ref{nonregressed2} (a)
a few very strong deviations from the bulk of the eigenvalues are seen, most significantly one real eigenvalue $\lambda_1\approx{}4.6$
and a conjugate pair of complex eigenvalues. 
Fig. \ref{nonregressed2} (b) is a detail of (a) where a clear shift of the bulk 
of the eigenvalues with respect to 
the Gaussian regime (circle) is observed. 
This shift can be attributed to two effects:
First, each deviating positive real eigenvalue $\tilde{\lambda}_i$ is
associated with a shift $s$ of the 'bulk' spectrum of $s \approx{}-\mathrm{Re(\tilde{\lambda}_i)}/N$ in
direction of the negative real axis. ('Departing' eigenvalues are those which have real parts larger 
than the radius of the theoretical support.)
The shift of the 'disc' pertaining to this effect is then the sum of all effects from departing eigenvalues, 
$ s_{tot}=-\frac1N \sum_{\tilde{\lambda}_i}\mathrm{Re}(\tilde{\lambda}_i) \approx{}-0.031$. 
A second contribution of the shift is due to the non-zero diagonal entries of the correlation matrices ${\bf C_1}$. 
The shift of the center of the disk explainable by the mean of the diagonal elements 
is $\bar C_1^{ii}=-0.029$, such that the overall displacement is 
$d=s_{tot}+\bar C_{ii}= - 0.060$. When corrected for the total shift we arrive at 
Fig. \ref{nonregressed2} (c).
We repeated the same procedure for ${\bf {C_1^{res}}}$, getting    
$d^{\rm res}=s_{tot}+\bar C_{ii}=-0.020-0.061= 0.081$; 
the resulting displacement corrected distribution is depicted in Fig. \ref{nonregressed2} (d).
The  shift of the center of the support is thus  quite simply explained.
\begin{center}
\begin{figure}
\begin{center}
\resizebox{0.5\textwidth}{!}{\includegraphics{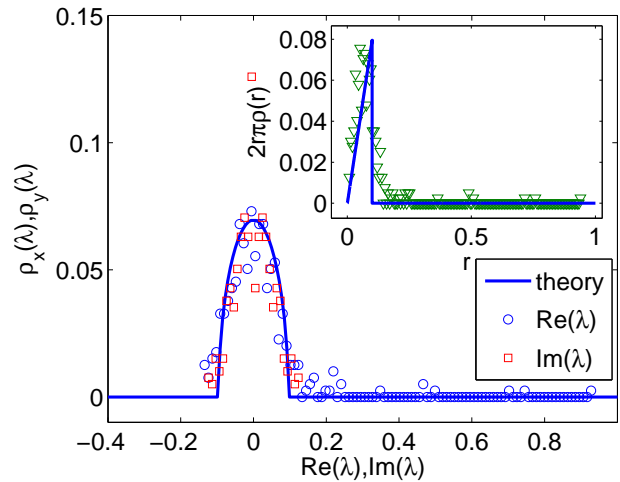}}
\end{center}
\caption{Projection of the empirical spectrum pertaining to
Fig. \ref{nonregressed2}c on the real and imaginary axis. The blue line is
the
analytical solution discussed in Section \ref{sec2}. The inset shows the empirical
distribution of $\rho(|\lambda|)$ compared with the analytical analogue $2r\pi\rho(r)$.
}
\label{againsttheory} 
\end{figure}
\end{center}

The eigenvalues lying outside the random regime should now be clearly associated with
specific non-random structures which will be examined below. For the eigenvalues
within the circle -- i.e. for the eigenvalues within the regime of Gaussian
randomness -- the natural
expectation would be that these follow the Gaussian predictions developed in
Section \ref{sec2}. 

In Fig. \ref{againsttheory} we compare predictions from Section \ref{sec2} 
with the empirical data , showing projections of empirical eigenvalue data onto the real and imaginary
axis. The inset shows the theoretical prediction of the radial density integrated over the complex plane,
$2r\pi\rho(r)$, compared with the empirical data, $\rho(|\lambda|)$. We 
chose a 'accumulated' representation since data quality would be
unsatisfying otherwise. The empirical spectra are truncated at
$\mathrm{Re}(\lambda)=1$. Given the modest eigenvalue statistics
($N_\lambda=400$) and the strong deviations outside the theoretical support,
the agreement between the theoretical predictions for Gaussian noise and the
empirical data seems rather satisfying. 

\subsubsection{Interpretation of deviating eigenvalues}

Strong deviations from the theoretical pure random prediction 
indicate significant correlation structure in the data. It is intuitively clear that 
eigenvalues departing positively (negatively) on the real axis with no or only a small
imaginary part will be the effect of symmetric (anti-) correlations. On the other
hand, complex conjugate eigenvalues departing on the imaginary axis will be attributable
to asymmetric, non-Gaussian correlations. 

Thus, the departures of the largest eigenvalue in Fig. \ref{nonregressed2} (a)
and (c) should be caused by a lagged correlation structure either 
pertaining to a group of stocks or to all of the stocks. 
On the other hand, we also see 
significant non-symmetric correlations in ${\bf X}$ reflected in complex-conjugate
pairs of eigenvalues with relatively large imaginary parts. 
The residuals ${\bf X^{res}}$ show a large negative real
eigenvalue indicating approximately symmetric anti-correlations between
stocks. Such a departure is not visible for ${\bf X}$.

For a closer inspection of which assets 'participate' in a given eigenvector
belonging to a deviating eigenvalue, 
one usually defines the inverse participation ratio for the eigenvectors $\vec{u}_i$,  
\begin{equation}
\mathrm{IPR}(\vec{u}_i) \equiv \sum_{\ell=1}^N | u_{i\ell} |^4  \quad.
\label{IPR}
\end{equation}
This ratio shows to which extent each of the $N=400$
assets contribute to the eigenvector $\vec u_{i}$. 
While a low  $\mathrm{IPR}$ means that assets contribute equally, 
a large $\mathrm{IPR}$ signals that only a few assets dominate the eigenvector.

\begin{figure}
\begin{center}
\resizebox{0.4\textwidth}{!}{\includegraphics{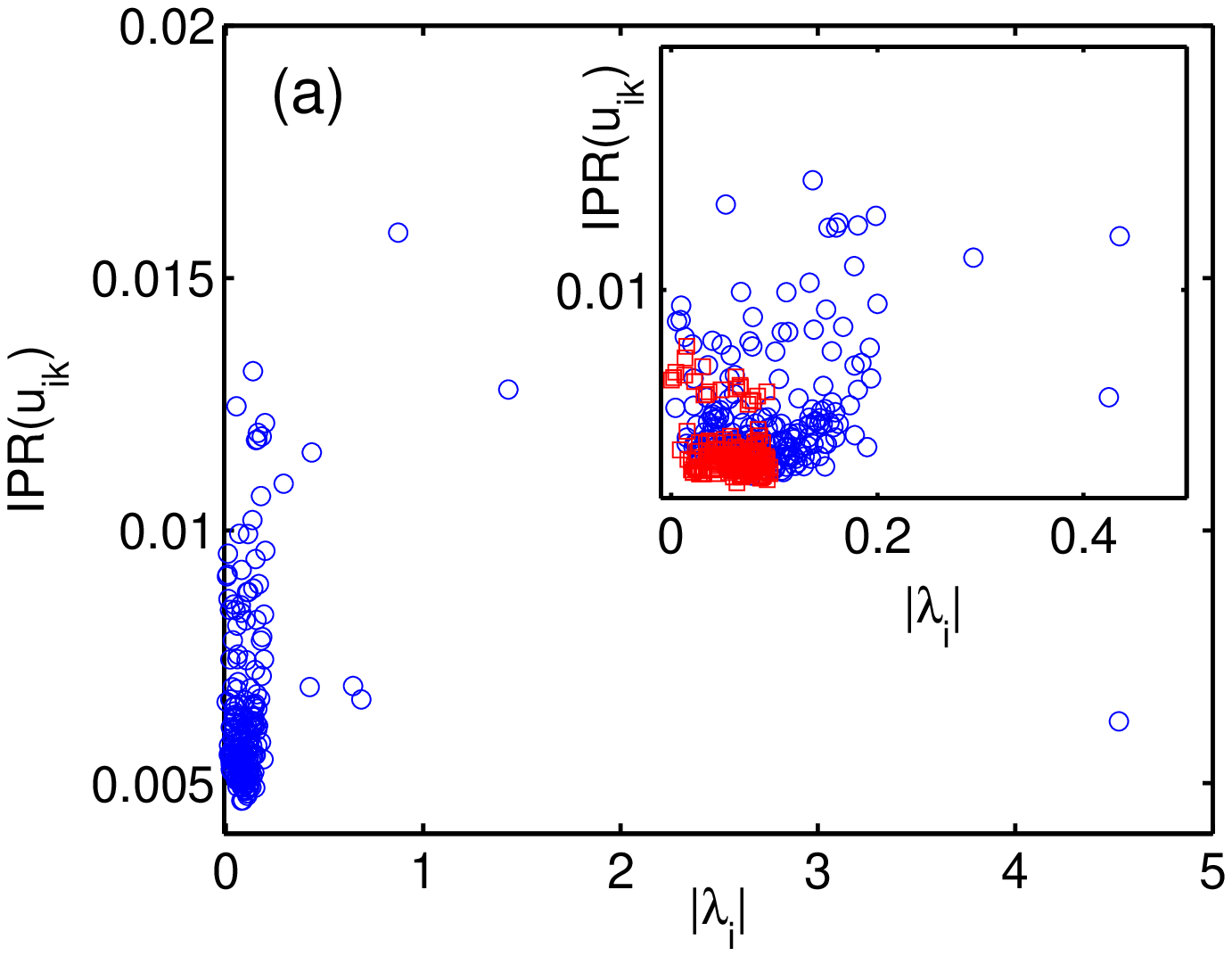}}\\
\resizebox{0.4\textwidth}{!}{\includegraphics{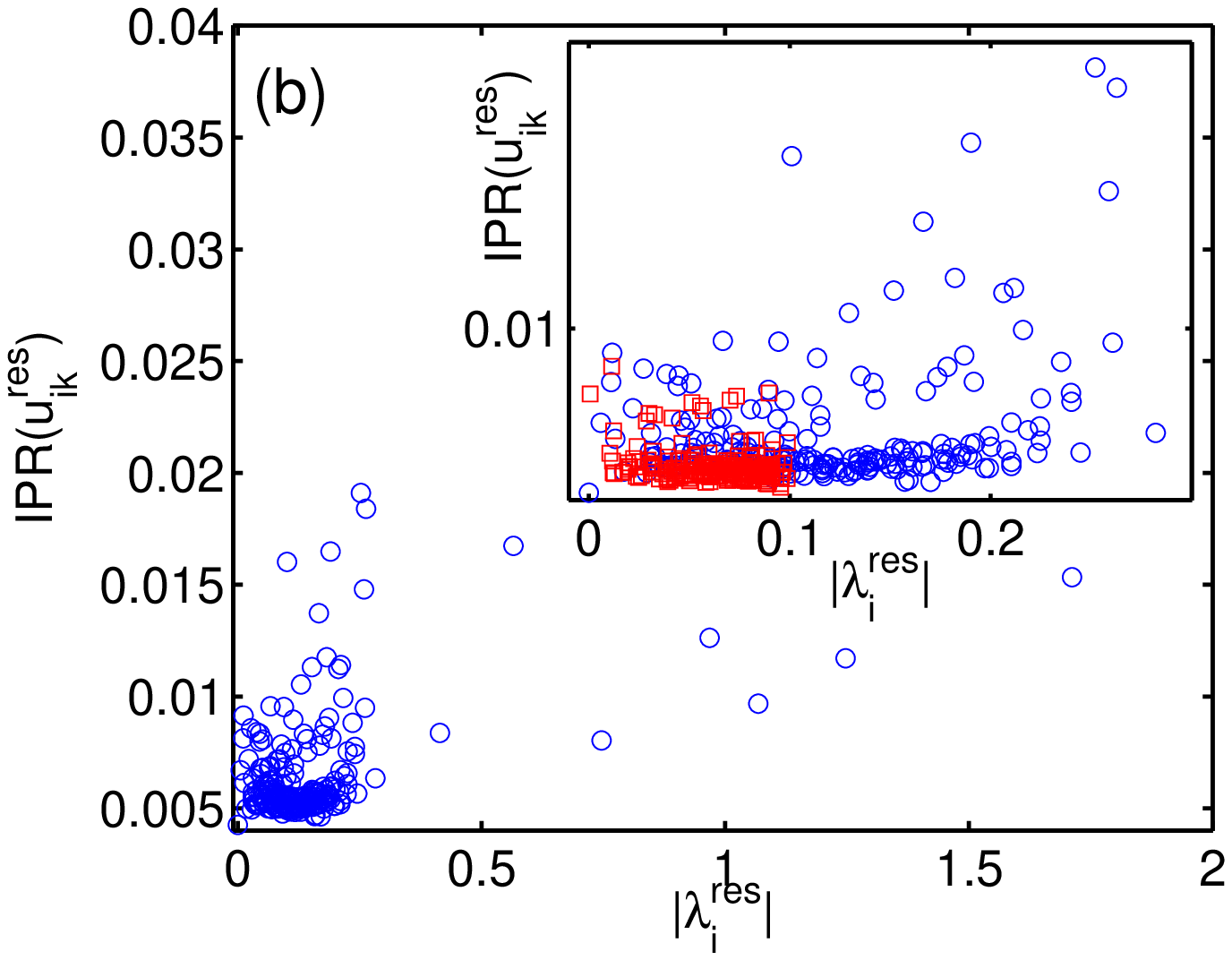}}\\
\end{center}
\caption{(a) Inverse participation ratio as defined in Eq. (\ref{IPR}) 
as a function of the absolute value of $\lambda_i$. Circles represent data 
from the empirical matrix, squares (inset) data from a random analogue,
obtained from iid gaussian distributed ${\bf X}$. 
(b) The same as above but for eigenvectors obtained from the data 
with the market mode subtracted out.}
\label{picipr}
\end{figure}

Figure \ref{picipr} (a) shows the IPRs for the empirical correlation matrix ${\bf C_1}$. 
The inset is a detail and also exhibits the IPRs from scrambled data 
(squares). It appears, that the 'random' regime is not confined to an 
approximately constant region of IPRs but varies quite widely. This is in contrast 
to the symmetric case where one has a constant IPR  for  eigenvalues stemming
from Gaussian randomness. 
We checked that the fluctuations observed here are already present in the Ginibre ensemble 
of real random asymmetric matrices and are thus not associated to
the specific structure of ${\bf C_{\tau}}$. 
It is clear, that the IPRs belonging to the random case not being bound to a line hinders 
the identification of the eigenvectors with 
strong influence of only a few components to a certain extent. 
However, one can nonetheless see that the 
largest departing eigenvalue $\lambda_1$ is characterized by a rather small
IPR, indicating an influence of a large number of assets. 
In contrast, some other deviant eigenvalues lie well above
the random regime indicating the influence of only few stocks.

Again, we compare with the situation found for the residuals ${\bf X^{res}}$
which is given in Fig. \ref{picipr} (b). 
On average, the IPRs of the deviating eigenvalues are larger than in (a),  
indicating a more clustered structure. 
We further analyzed and IPR-like quantity only based on the imaginary parts,  
$\mathrm{IPR}(\mathrm{Im}(\vec{u}_i))=\sum_{\ell=1}^N\mathrm{Im}(u_{i \ell})^4$, 
and found pure random behavior, except for $\lambda_2=\lambda_3^*$ (not shown).
With evidence  at hand for some group structure in the lagged-correlations, 
we now take a closer look at these structures.

\subsubsection{Sector organization in time-lagged data}

\begin{center}
\begin{figure}
\begin{center}
\begin{tabular}{cc}
{\Large original} & {\Large market removed} \\
\resizebox{0.25\textwidth}{!}{\includegraphics{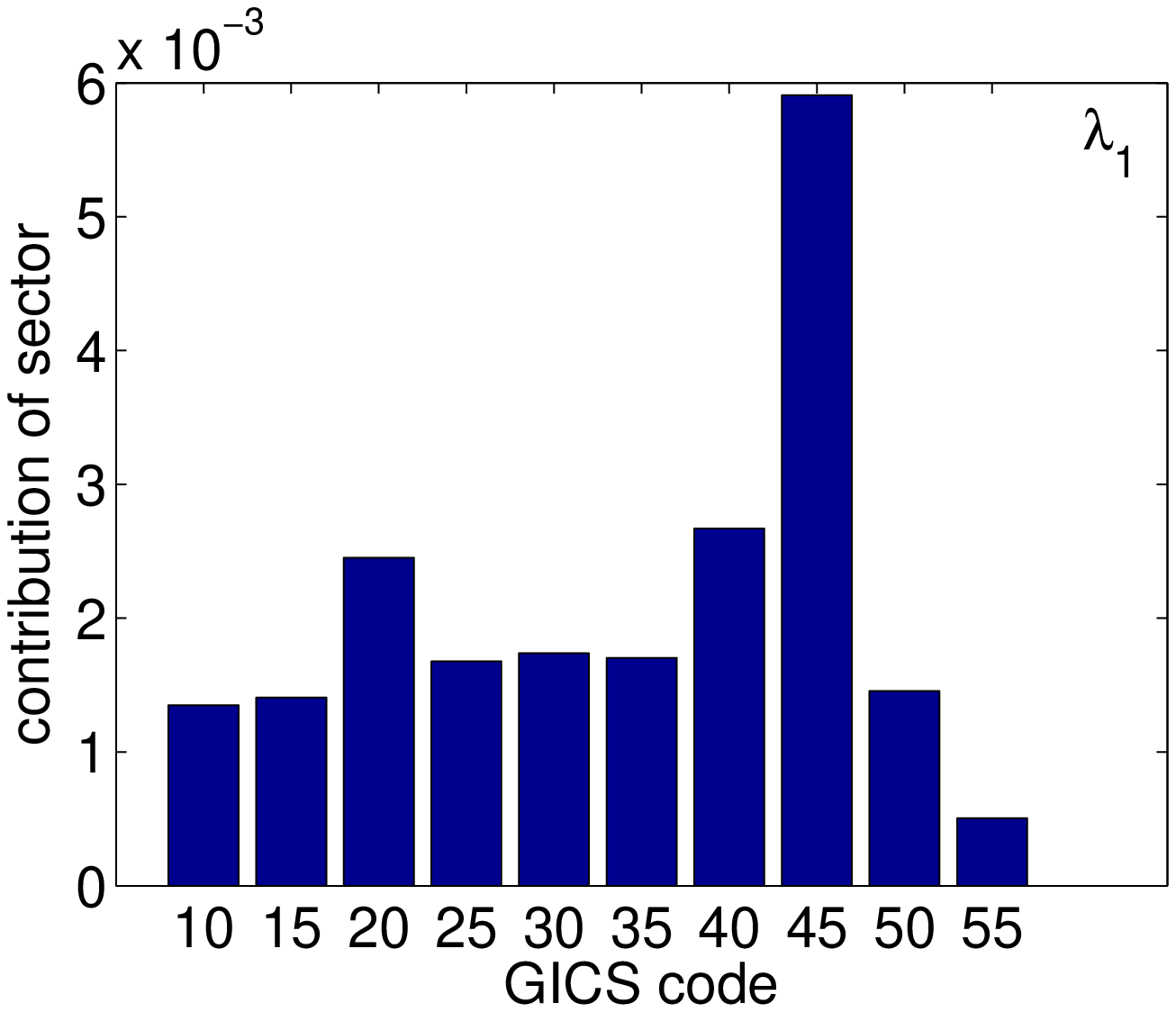}  } & \resizebox{0.25\textwidth}{!}{\includegraphics{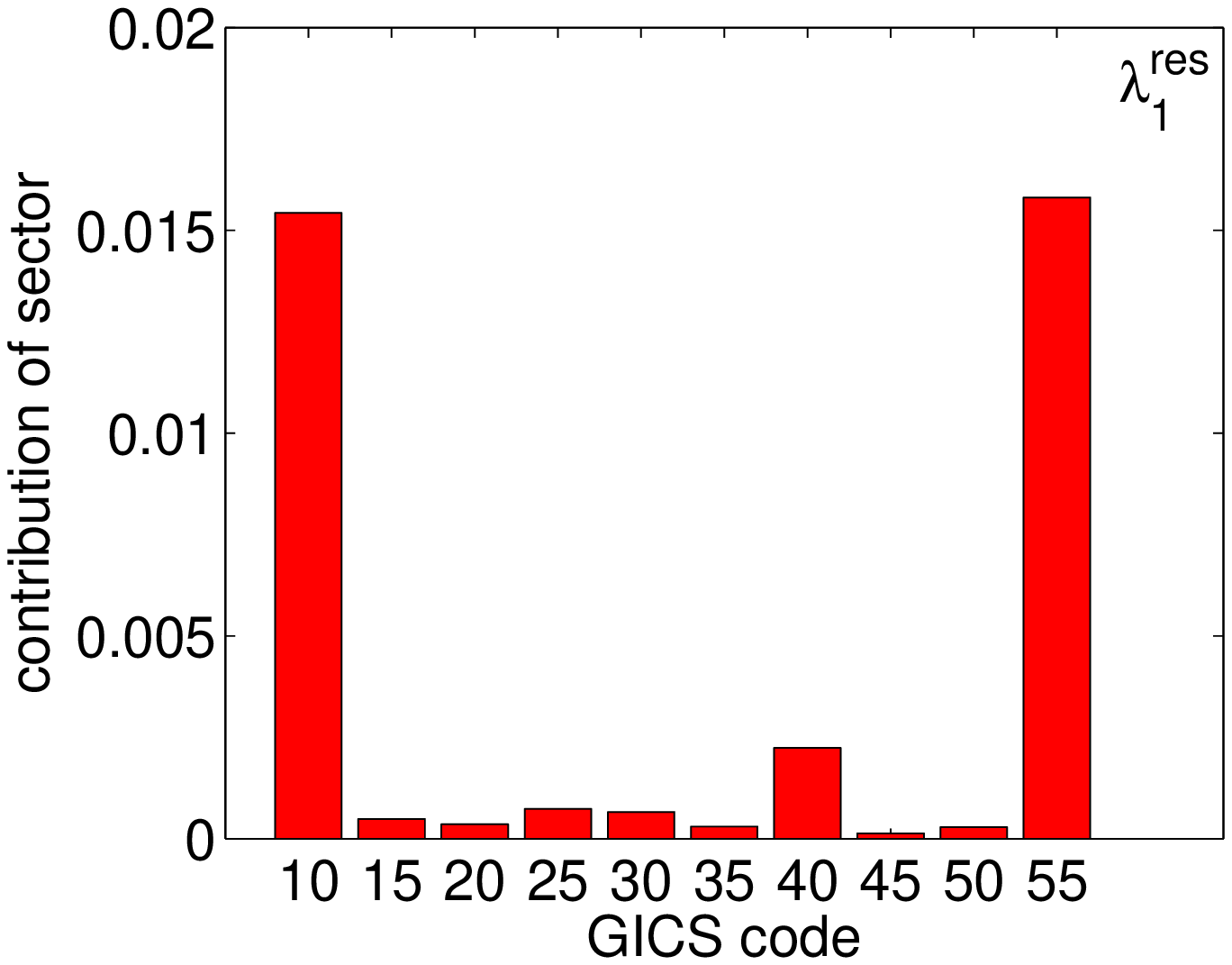}} \\
\resizebox{0.25\textwidth}{!}{\includegraphics{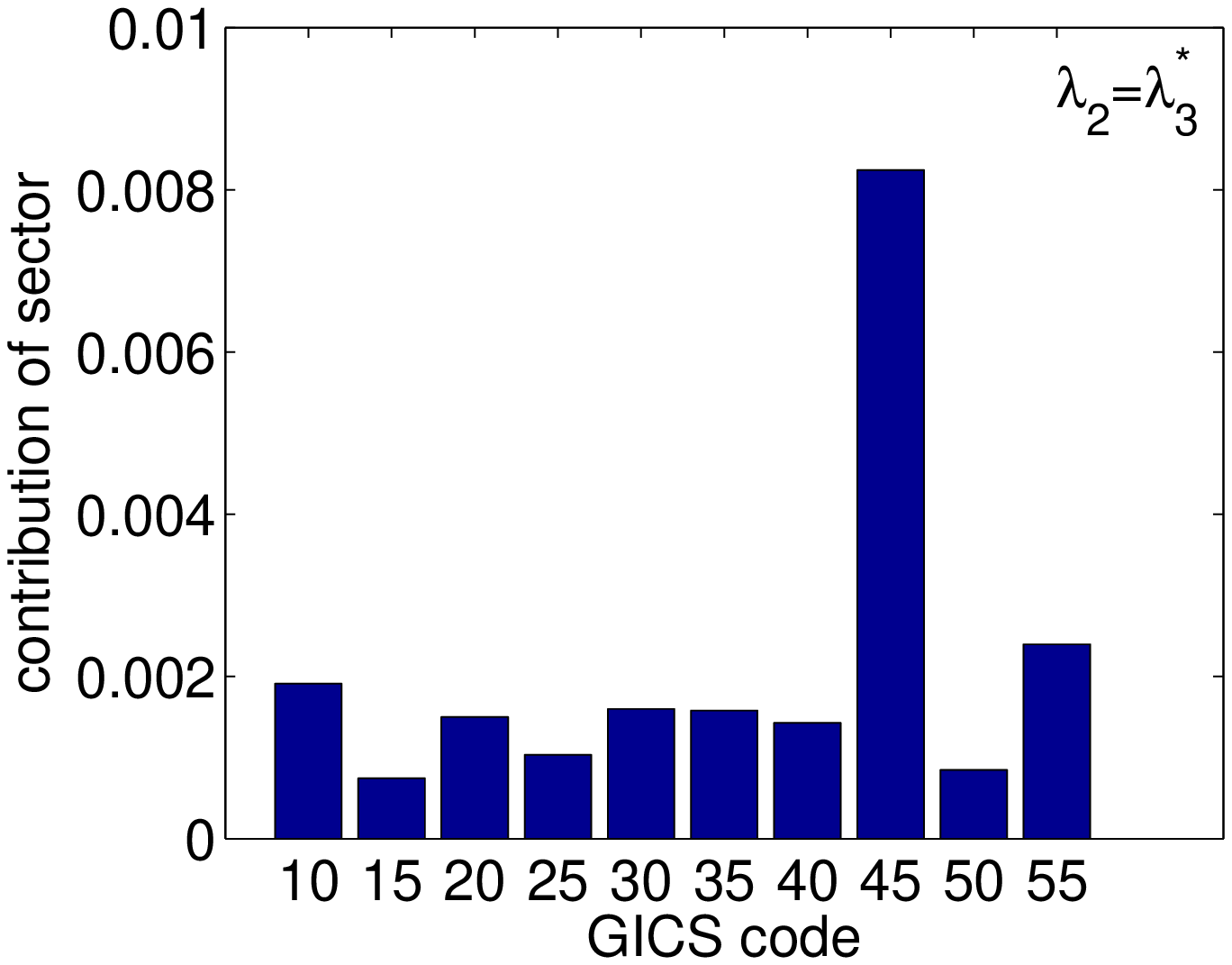} } & \resizebox{0.25\textwidth}{!}{\includegraphics{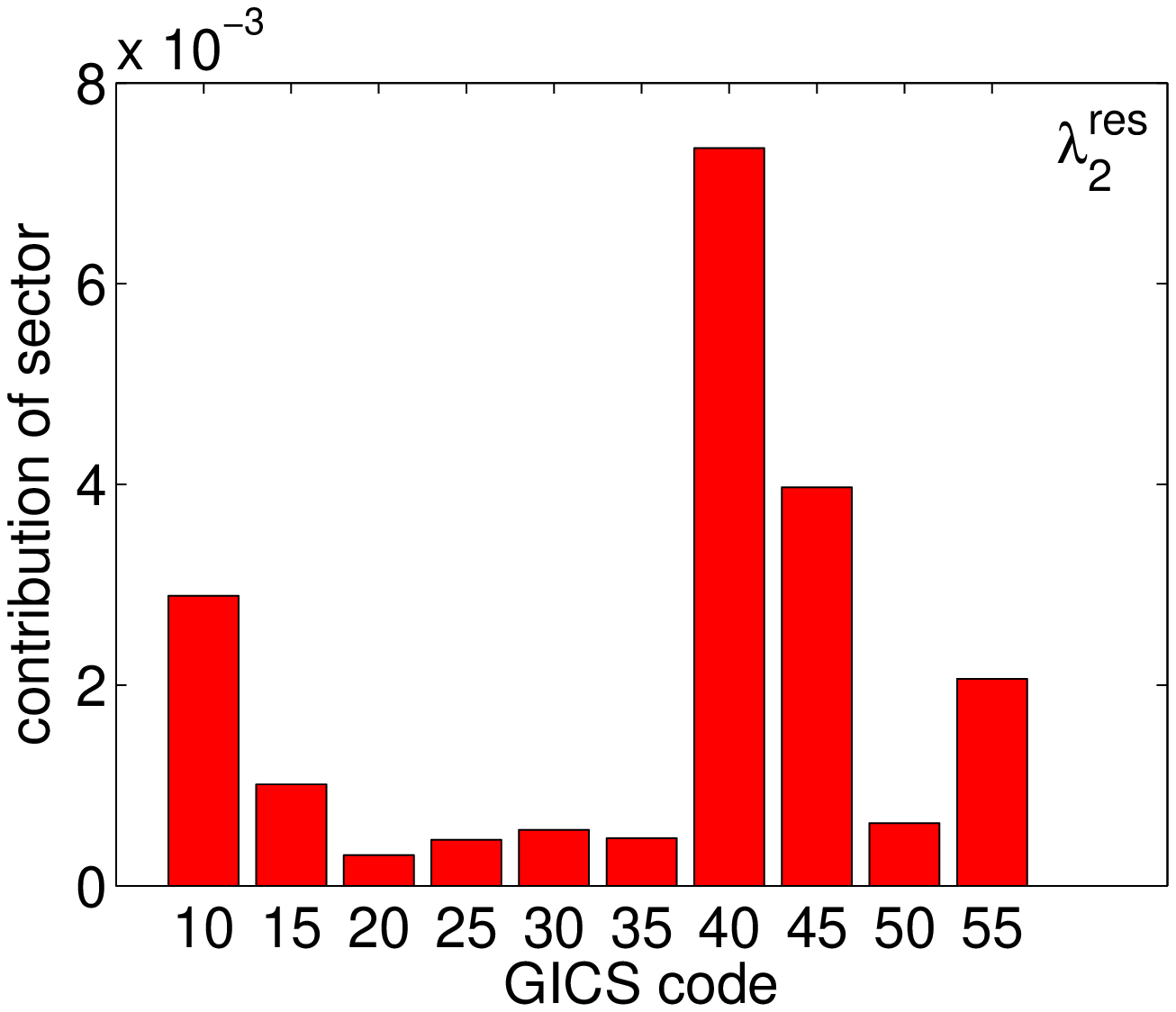}} \\
\resizebox{0.25\textwidth}{!}{\includegraphics{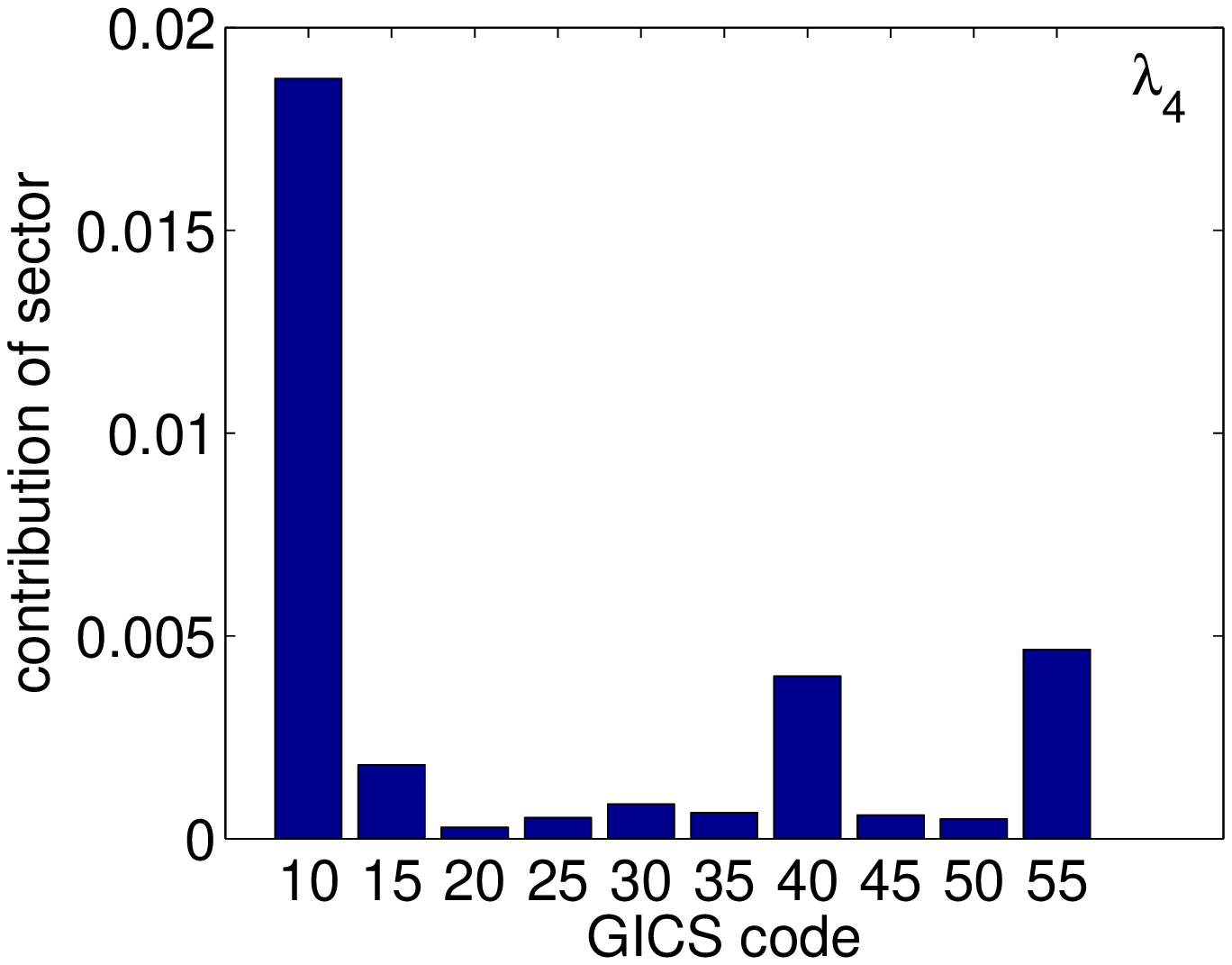}  } & \resizebox{0.25\textwidth}{!}{\includegraphics{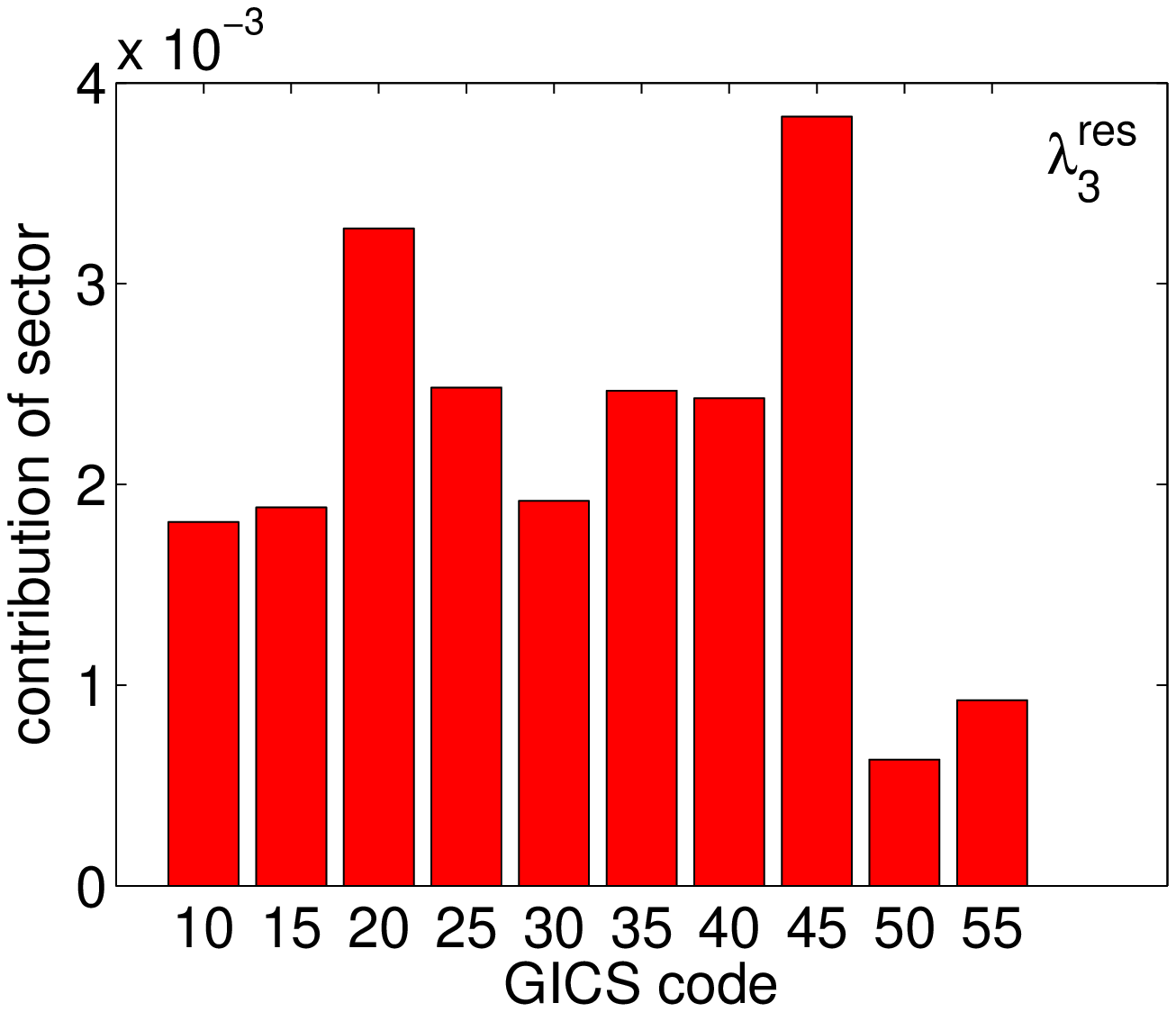}} \\
\resizebox{0.25\textwidth}{!}{\includegraphics{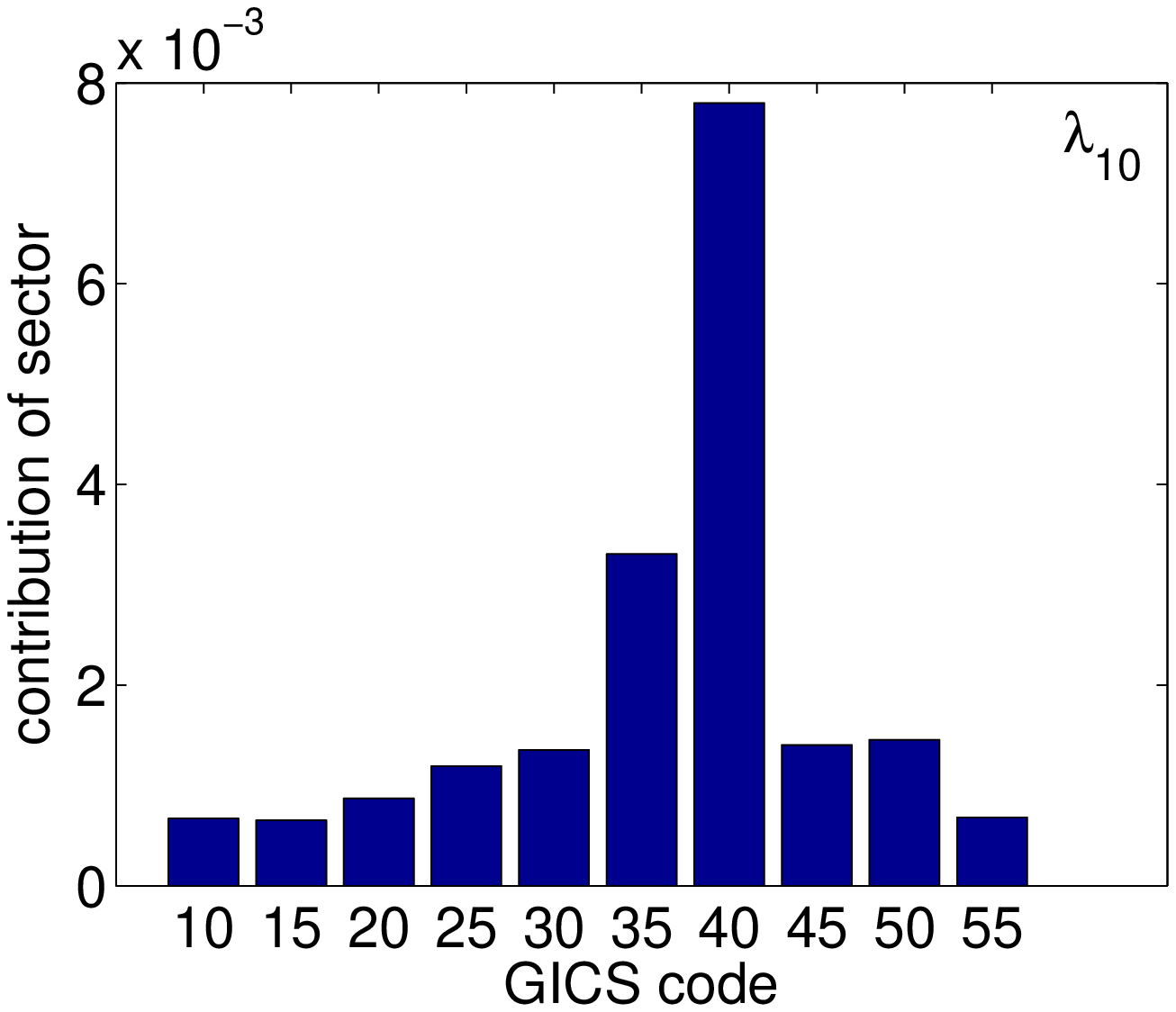} } & \resizebox{0.25\textwidth}{!}{\includegraphics{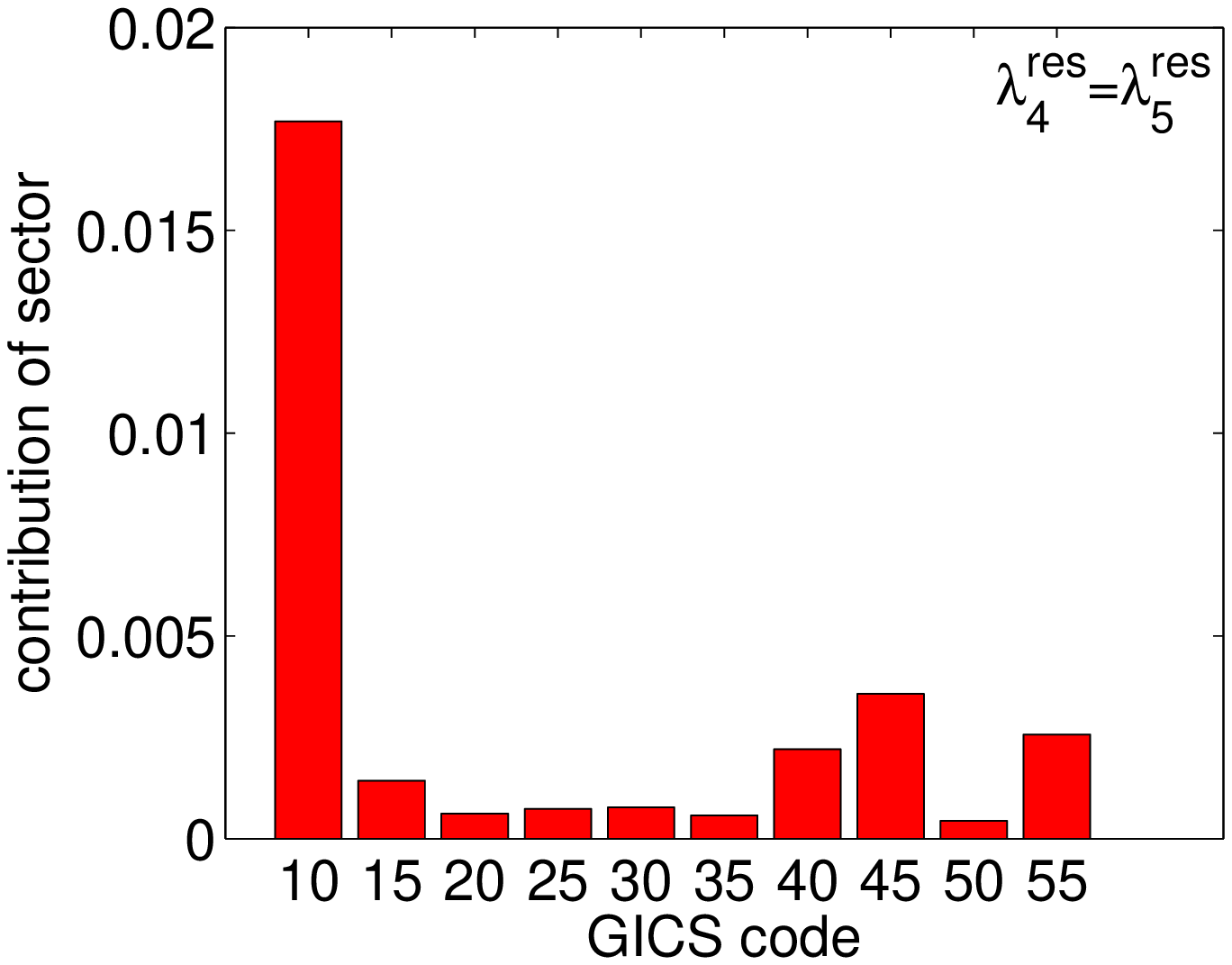}}
\end{tabular}
\end{center}
\caption{
\label{participation}
Strength of participation, $I_{si}$, of the ten main sectors of the S\&P500
(according to the GICS code) to eigenvectors $\vec{u}_i$
for some selected eigenvalues $\lambda_i$.}
\end{figure}
\end{center}

It is  well known from RMT applications to covariance matrices ($\tau=0$) of
financial data, that the eigenvectors $\vec{u}_i$ 
of large eigenvalues can be associated with the sector organization of  
markets. 
Let us label the different sectors with $s$, and define 
\begin{equation}
 \Delta_{sk}=\left\{
 \begin{array}{ll}
   1 & \textrm{if stock $k$ belongs to sector $s$}\\
   0 & \textrm{otherwise}\\
 \end{array}
\right.
\quad .
\end{equation}
To visualize the influence of each sector $s$ to a given eigenvector $i$, 
we calculate
\begin{equation}
  I_{si}  \equiv \frac{1}{N_s}\sum_{k=1}^N{\Delta_{sk}|u_{ik}|^2} \quad , 
\label{xsi}
\end{equation}
where $N_s$ is the number of stocks in the respective sector, $s$. 
We evaluate Eq. (\ref{xsi}) 
for the S\&P500, using the standard sector classification scheme,
the so-called GICS code, which is summarized in Table \ref{GICS}. 
\begin{table}
\begin{tabular}{lcc}
\hline
\hline
Sector & GICS & No. of Stocks $N_s$\\
\hline
\hline
Energy                 & 10  & 22\\
Materials              & 15  & 27\\
Industrials            & 20  & 44\\
Consumer Discretionary & 25  & 63\\
Consumer Staples       & 30  & 35\\
Healthcare             & 35  & 40\\
Financials             & 40  & 71\\
Information Technology & 45  & 63\\
Telecommunication      & 50  & 11\\
Utilities              & 55  & 24\\
\hline
\end{tabular}
\caption{\label{GICS} 
Global Industry Classification Standard (GICS code), 
for the 10 main sectors of the S\&P500 with the number of stocks in these sectors,  
see \emph{www.standardandpoors.com}. }
\end{table}
Figure \ref{participation} shows the contributions of the sectors to a set of 
selected  eigenvalues for the original (left column) and the market-mode 
removed data (right column).
In the case of the original data, the information technology sector seems to play a decisive role 
for the largest 3 eigenvalues, namely $\lambda_1$ and $\lambda_2=\lambda_3^*$. 
This sector thus explains a large part of the most distinctive non-random (symmetric and
asymmetric) structure in ${\bf C_1}$.
For other eigenvectors, as for example $\lambda_4$ and $\lambda_{10}$ 
and others not shown here, a distinctive role is played by the energy and financial sector, respectively.
Results for ${\bf C_1^{res}}$ (right column) also show remarkable 
deviations from the Gaussian efficient market prediction (equal contribution of the individual
sectors). Here, the largest eigenvalue $\lambda_1^{\rm res}$ is associated with a
strong participation of the energy and utility sectors. In the
second eigenvector, the financial sector is dominant, whereas the
eigenvalue associated with the strong negative departure on the real axis,
$\lambda_3\approx{}-1$, is not dominantly influenced by any sector. 
For $\lambda_4=\lambda_5^*$ we find a strong influence of the energy
sector. Other eigenvectors also indicate a strong sectorial contribution (not shown).

\begin{center}
\begin{figure}
\begin{center}
\resizebox{0.5\textwidth}{!}{\includegraphics{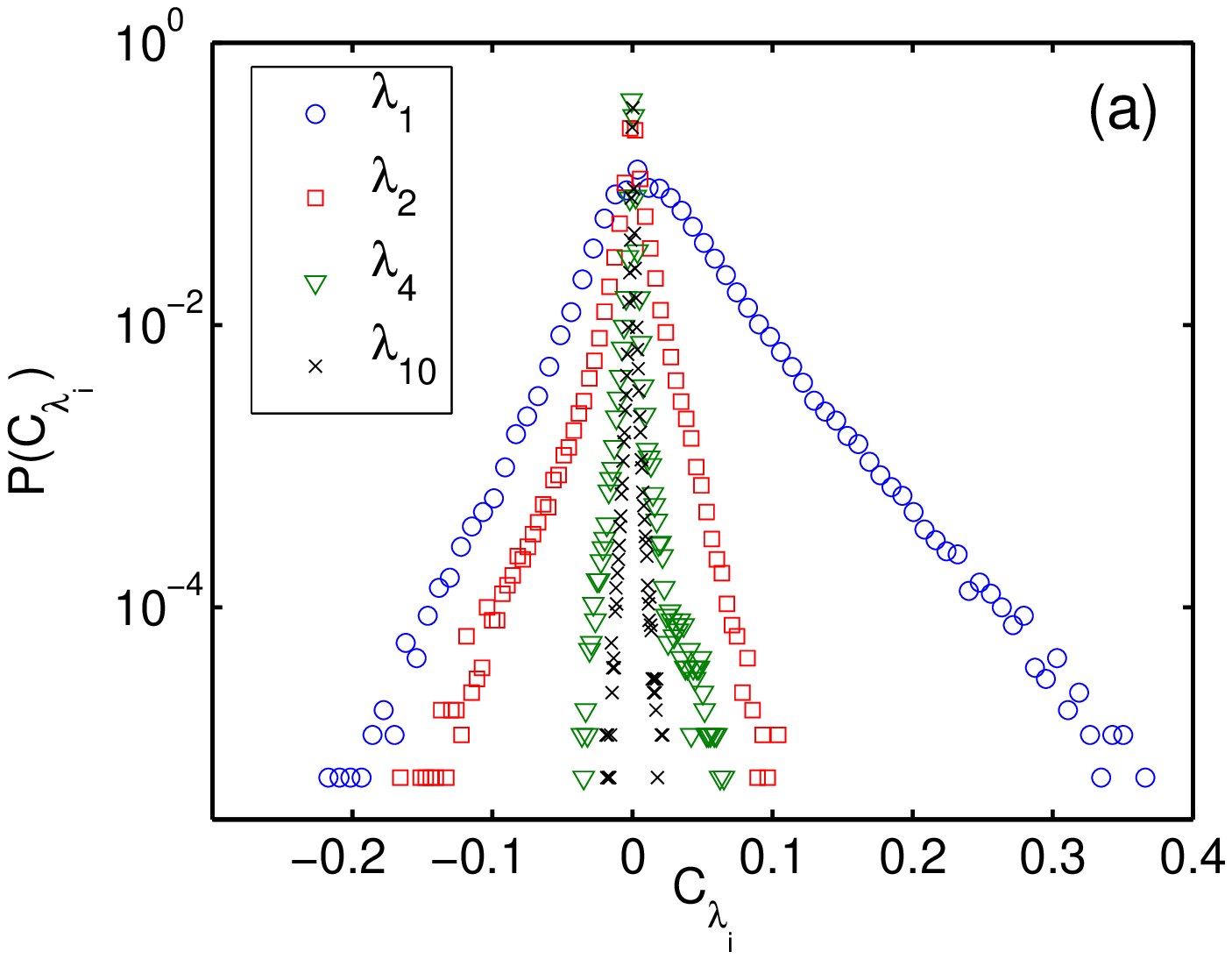}}
\resizebox{0.5\textwidth}{!}{\includegraphics{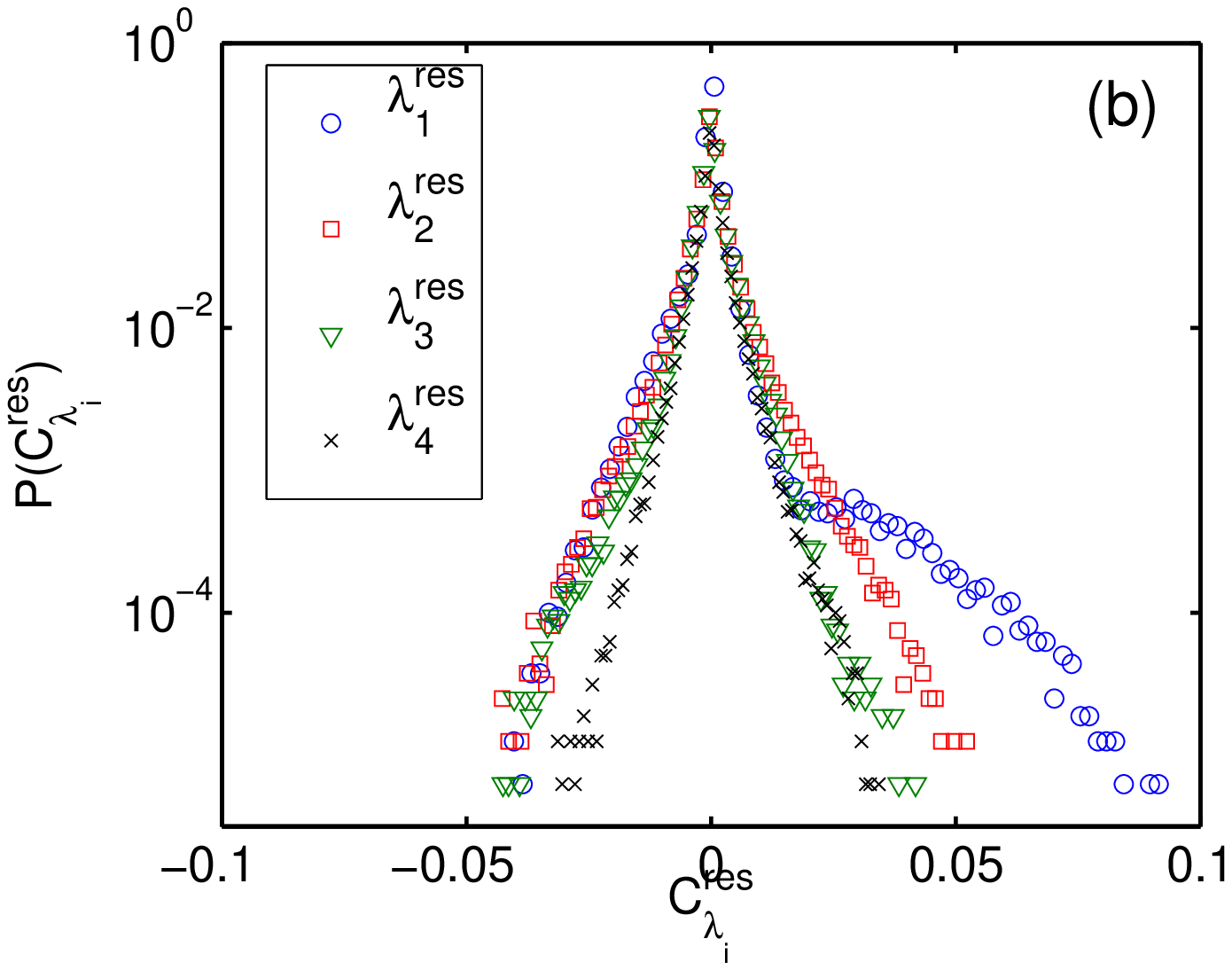}}
\end{center}
\caption{\label{lhist_sup} Histograms of entries in ${\bf C_1}^{\lambda_i}$ 
for several strongly deviating  eigenvalues, for original (a), and  
market-mode removed data (b).}
\end{figure}
\end{center}

For a quantitative discussion of the structure imposed by  the individual
eigenvectors and eigenvalues we decomposed the (square) correlation matrices with respect to 
individual eigenvalues, 
\begin{equation}
\label{decomposition}
{\bf C}_{\lambda_i} = u_{ij}\mathrm{diag}
(\lambda_i)u_{ij}^{-1}\quad , 
\end{equation}
where $\mathrm{diag}(\lambda_i)$ denotes a diagonal matrix with only one entry
at the respective position, associated with eigenvalue $\lambda_i$. 
In Fig. \ref{lhist_sup} (a) we display histograms of the elements of ${\bf C_1}^{\lambda_i}$
in the same way as in Fig. (\ref{pic22}).
The largest contribution to ${\bf C_1}$ is seen to originate from $\lambda_1$, 
and  tails seem to follow a distinctive exponential  distribution. 
Thus, the structure associated with $\lambda_1$ is definitely 
not Gaussian and exhibits specific (exponential) behavior which is not visible in the
distributions of the elements of the full matrix ${\bf C_1}$. 
The complex pair $\lambda_2=\lambda_3^*$ carries predominantly
negative correlations. 
The following eigenvalues contribute much less.  The 'humps' in
the histograms, e.g. seen for $\lambda_2=\lambda_3^*$ and $\lambda_4$, indicate some deterministic structure.
In Fig. \ref{lhist_sup} (b) the same is shown for the  market removed data. 
The positive tails of the distribution of the entries of ${\bf C_{1}}^{\lambda_1^{\rm res}}$
strongly deviate from the Gaussian regime. This 'hump' can be understood as a
consequence of strong correlations of sectors 10 and 55, seen in Fig. \ref{participation}. 
This effect is also visible in a network visualization of the market removed matrix.
We will now proceed to such a network view to visualize and further discuss the findings of strong sectorial 
contribution and strongly anomalous distributions ${\bf C}_{\lambda_i}$.

\subsubsection{Lead-lag networks}

\begin{center}
\begin{figure}
\begin{center}
\resizebox{0.42\textwidth}{!}{\includegraphics{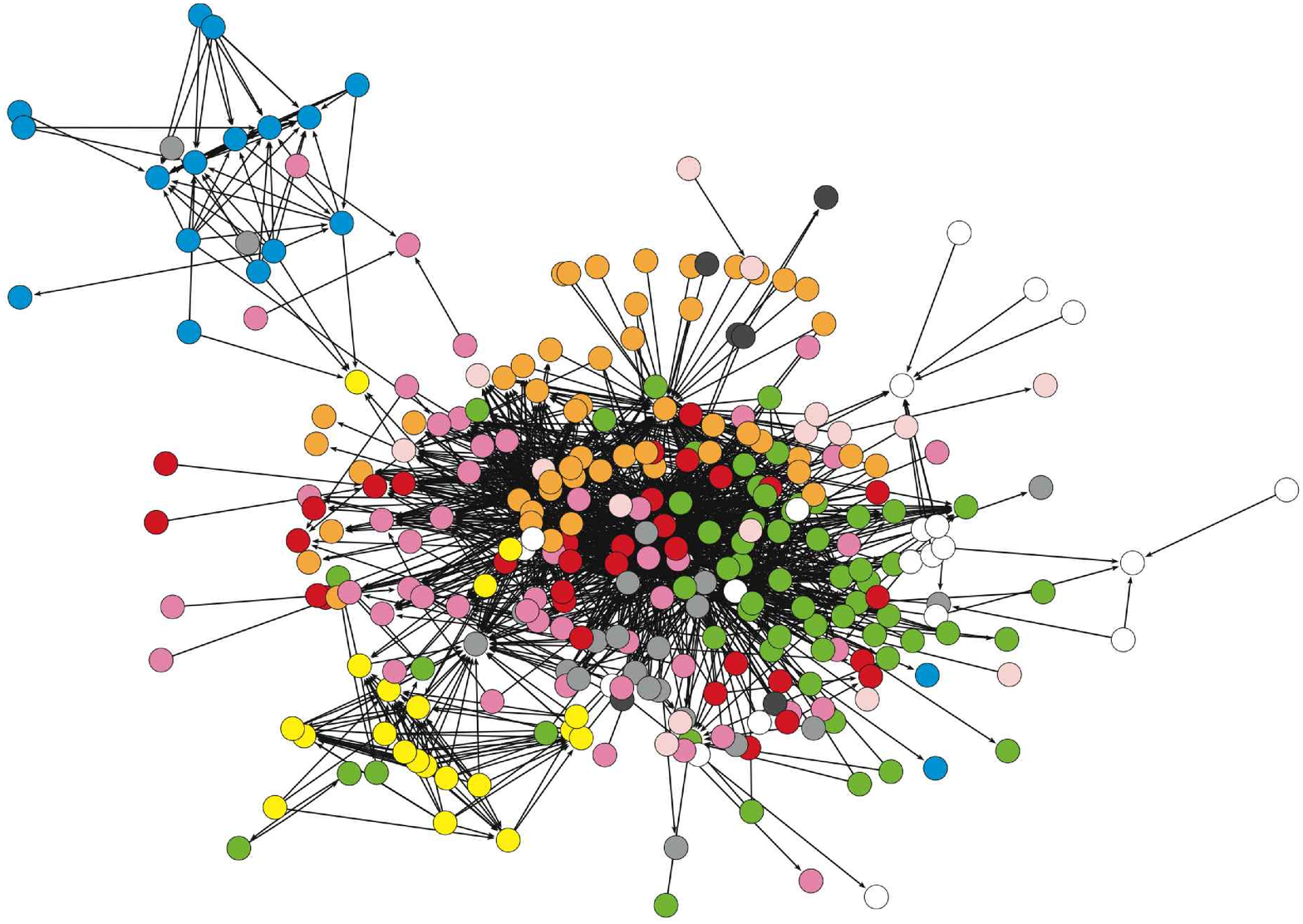}} \\
\hspace{3cm} {\Large (a)} \\
\resizebox{0.42\textwidth}{!}{\includegraphics{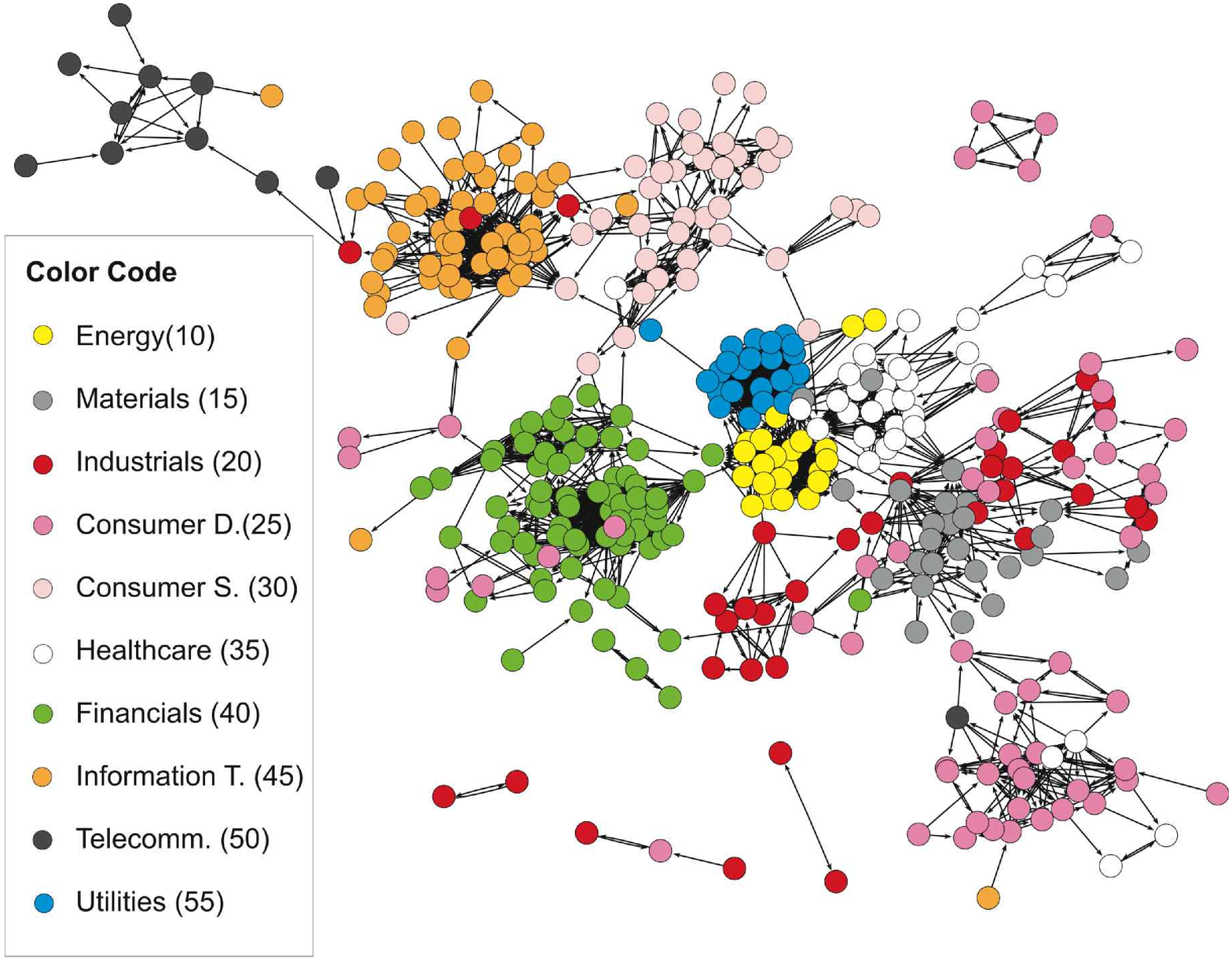}} 
\\
\hspace{3cm} {\Large (b)} \\
\resizebox{0.42\textwidth}{!}{\includegraphics{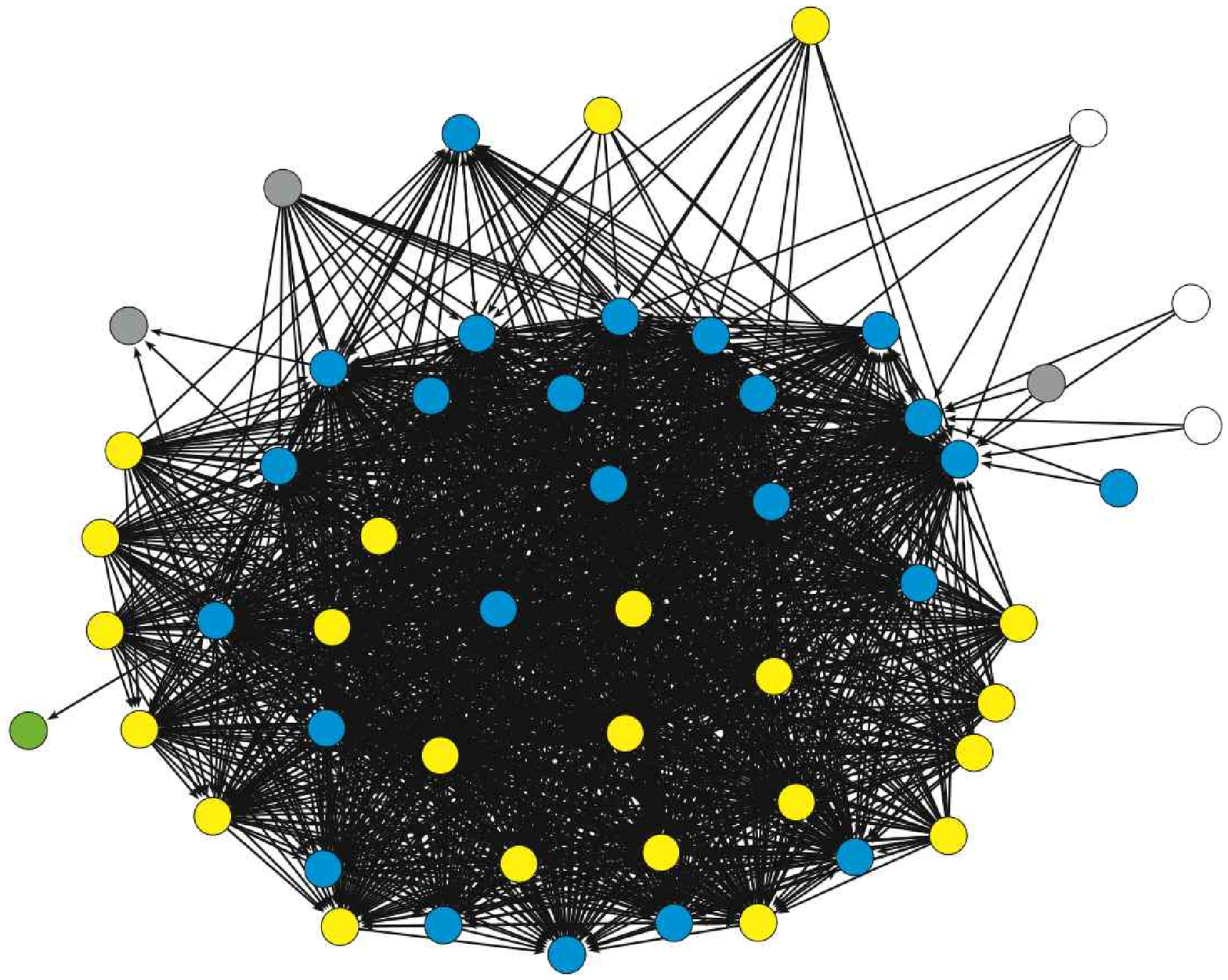}} \\
\hspace{3cm} {\Large (c)}

\end{center}
\caption{Network view of ${\bf C_1}$ (a). A Link was drawn for 
$C_1^{ij}>0.09$.
The situation for the regressed scenario is shown in (b) with a 
threshold of $C_1^{ij \, {\rm res}}>0.033$. (c) Shows the correlation 
network for stocks belonging to the largest eigenvalue in the regressed data (for $C_1^{ij \, \lambda_1}>0.13$). 
Two sectors (Energy and Utilities) are tightly bound together.
All network pictures are results from a Kamada-Kawai algorithm.}
\label{nw1}
\end{figure}
\end{center}

Comparing eigenvalue spectra of the residuals with those  of the initial 
data (Figure \ref{nonregressed2}), it is apparent that the market mode has a 
clear influence on the deviations and that the largest eigenvalue for the 
residuals is significantly reduced.
As a matter of fact, one would expect that removing the (equal-time) market-mode also
eliminates much of the correlations pertaining to small firms  
driven by large companies or similar 'star-like' structures (i.e. any network
structure where one stock leads or lags many other stocks).
In Fig. \ref{nw1} (a) we show a network view  of the ${\bf C_1}$ 
correlation matrix, 
where a link is drawn for any $C_1^{ij}>0.09$; (b) is the same after 
removal of the
market mode, and $C_1^{ij \,\, {\rm res}}>0.033$. Clearly,  while in (a) there 
is not much clustering (except maybe for the utility sector), in the market 
removed scenario distinctive clustering appears. 
As in the previous section, we identified the nodes with 
the 10 most important sectors in the market. Nodes are colored according to 
these sectors in Fig. \ref{nw1} along the lines of the accompanying color scheme.
The identified clusters correspond very nicely with industry sectors, as
was found quite some time ago for $\tau=0$. 

Returning to an analysis of the original data, 
we look at networks derived from individual
matrices ${\bf C}_1^{\lambda_i}$, Eq. (\ref{decomposition}), 
to visualize some 'qualitative structure' associated with strongly
deviant eigenvalues and thus associated to the most 'orthogonal' aspects of
overall-deviations.

For the largest eigenvalue $\lambda_1$, we investigate a few assets from the
Information Technology (IT) sector leading stocks of different sectors (not
shown) with positive lagged correlations. 
The most pronounced hubs from IT were found to be
AMAT, BRCM, INTC, KLAC, LLTC, MSFT, MXIM, NVLS, YHOO and XLNX. Quite
similarly, the most
prominent features of the conjugate pair $\lambda_2=\lambda_3^*$ can be
associated with a hub-like influence of the IT sector -- this time, however, with a
negative lagged correlation. Networks pertaining to $\lambda_4$ and
$\lambda_{10}$ primarily exhibited intersectorial ties of the Energy and
Financial sector, where we also observed hub-like anti-correlations 
pointing from stocks of the Financial sector to the Energy sector.

For the lagged correlation matrix of the residuals ${\bf X^{res}}$, the 
largest eigenvalue $\lambda_1^{\rm res}$ shows a strong clustering 
of Energy \& Utility sector, which is shown in Fig. \ref{nw1} (c).
The fact that practically no assets apart from the Energy
and Utilities sector are represented is fully conforming with the top right panel of Fig. 
\ref{participation}. 
The tight binding of these sectors is also seen in Figs.  \ref{nw1} (c),
and \ref{lhist_sup} (b). In the latter, the strong tail corresponding to 
positive correlations of $\lambda_1^{\rm res}$ seems to be a consequence of this binding.
The second largest eigenvalue, $\lambda_2^{\rm res}$, demonstrates organization of the 
Financial sector where some stocks -- namely BAC (Bank of America), FITB (Fifth Third Bank) and  C 
(Citigroup Inc.) -- dominate the others (not shown).
Closer inspection of the negative eigenvalue 
$\mathrm{Re}({\lambda_3})\approx{}-1$
reveals, that it is mostly associated with time-lagged
anti-correlations between various  sectors; eigenvalue $\lambda_4=\lambda_5^*$ exhibits clustering of the Energy and 
the Consumer Staples sector.

In general, the analysis of the residuals effectively 
reveal secondary information not seen before, which is mainly attributable to the 
sectorization of stocks. Inferring from causes to effects, this fact may explain in part or all of the
well investigated equal-time cross-correlations, see e.g. 
\cite{oldlaces} for a short description of an adequate model. In contrast to
the residuals, the original data exhibits lots of hub-like interactions, where
the  assets lagging the hubs do not seem to belong to a specific sector.
The most pronounced leading hubs are stocks from the IT sector which has
apparently 'lead' the market within an observed time-period.
As a side comment, it does not seem to us that the associated leading stocks were the
ones with the highest market capitalization as would be implied by the finding of \cite{lo_kinlay}.

\section{Time Dependence}
\label{sec4}

\begin{figure}
\begin{center}
\resizebox{0.5\textwidth}{!}{\includegraphics{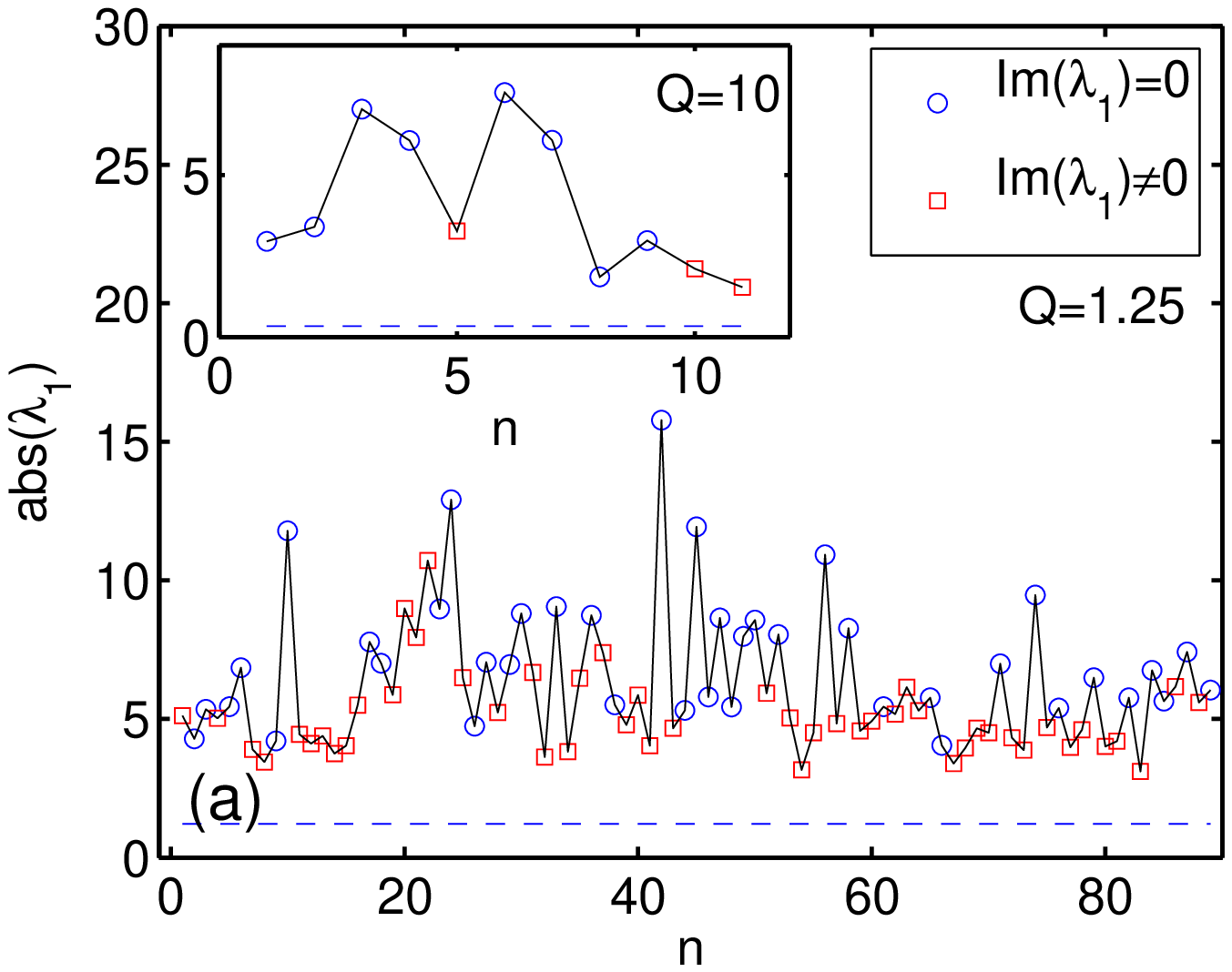}}
\resizebox{0.5\textwidth}{!}{\includegraphics{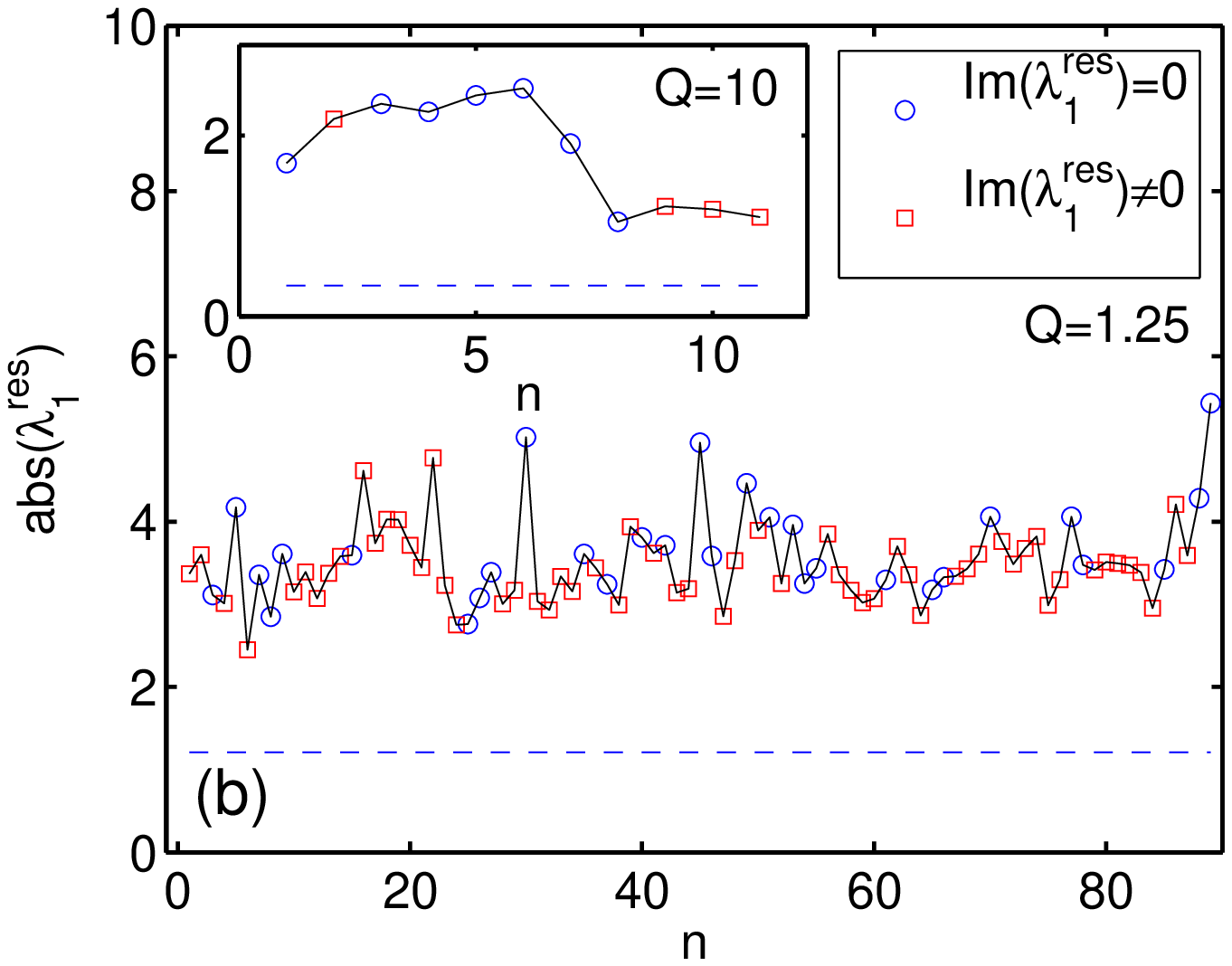}}
\end{center}
\caption{\label{timedep1} (a) Time dependence of the largest eigenvalue of ${\bf C}_1(T_n)$ as
a function of the period index $n$ for $T=500$ (main figure)
and $T=4000$ (inset). Values are plotted as blue circles if the largest
eigenvalue is located on the real axis ($\mathrm{Im}(\lambda^{max}_n)=0$) and as red
squares otherwise. (b) Same for ${\bf C_\tau^{res}}$.}
\end{figure}

In this section we discuss the time-dependence of the correlation matrices.  
We can immediately use the prediction of the support of the eigenvalue spectra in the complex plane
$\mathcal{C}$ to determine 
a minimum sampling period $T$ (or equivalently a minimum value of $Q$)
at which the estimated cross-correlations still exhibit non-random structure.
This is possible since we know that if eigenvalues are outside the support the data is 
non-random. Reducing $T$ too much one expects to arrive a very noisy estimate of the 
lagged correlation matrix,  
which will manifest itself in having no departing eigenvalues at all.

We calculate ${\bf C}_1{}(T_i)$ for consecutive, non-overlapping time periods
$T_i$ and find that --
very remarkably -- down to a information to noise ratio of $Q\approx{}1.25$,
clear deviations from the predicted support occur. 
This means that even though  noise is drastically increased for low values 
of $Q$, non-random structures prevail even at short time-scales.

More specifically, we analyzed 11 correlation matrices obtained from time slices of 
$4000$ observations (Q=10), and  89 matrices for $500$ time points each.
For each individual sub-period $T_n$, we compute 
lagged correlation matrices ${\bf C}(T_n)$ for the raw data as well as on the matrices resulting from the
regression model, ${\bf C^{res}}(T_n)$. Figure \ref{timedep1} (a) shows a plot of the
absolute value, $\mathrm{abs}(\lambda_n)$, of the maximal eigenvalue found for 
each sub-period, indexed by
$n$. The dashed blue line corresponds to the prediction of the support
$r_{max}$. We immediately recognize that for $Q=10$, as well as for $Q=1.25$
the largest eigenvalue lies significantly above the noise regime. On the other
hand, the absolute value of the largest eigenvalue is quite volatile and
anti-persistent for $Q=1.25$. We also observe that the 
largest eigenvalues with non-zero imaginary parts (red squares) mainly occur at low values of
$\mathrm{abs}(\lambda_n)$, whereas real eigenvalues occur at absolute values.
If the eigenvalue is real, the lead-lag network is dominated by strong,
approximately symmetric effects; for imaginary eigenvalues the network is
dominated by asymmetric correlations, i.e. anti-correlations may play a
distinctive part too.
 We find that if an eigenvalue $\lambda_1$ was real (i.e. marked
by a blue circle in Fig. \ref{timedep1}), the analysis of the preceding
sections always identified the IT sector mainly
contributing to $\vec{u}_1$ (for $Q=10$). On the other hand, if the largest
eigenvalue was imaginary, no unique interpretation appeared to be valid for
all of the sub-periods. 

In Fig. \ref{timedep1} (b) we show the same for our continuing antagonist ${\bf
X^{res}}$. Again, we observe $\mathrm{abs}(\lambda_n)$ being clearly located
above the random frontier for all sub-periods. The movement of
$\mathrm{abs}(\lambda_1)$ is less volatile. 
Closer investigation of the underlying eigenvalues for $Q=10$ revealed
changing participation of the sectors (measured by the quantity $I_{si}$ as
defined in Eq. (\ref{xsi})). In effect, for all of the 11 sub-periods
either the Energy (in periods 6-9) or the Utilities sector (in periods 3, 5) appeared as primarily
contributing. In the rest of the periods, both of these sectors were
represented strongly in $I_{si}$.

\begin{figure}
\begin{center}
\resizebox{0.5\textwidth}{!}{\includegraphics{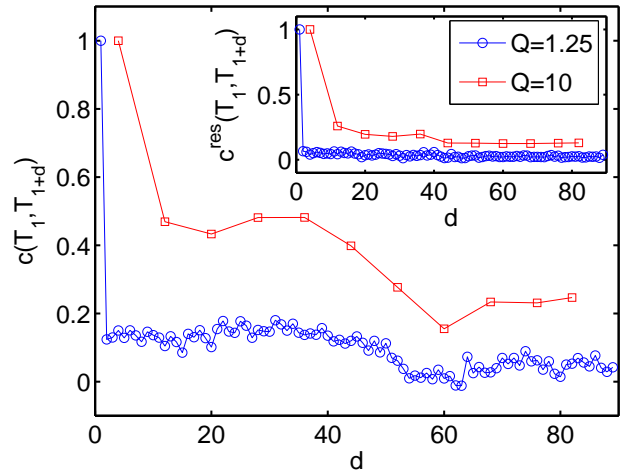}}
\end{center}
\caption{Matrix element correlation $c(T_1,T_{1+d})$ as described in Eq. (\ref{corr2formula}) for various time lags
$d{}$ for the original data and for the residuals ${\bf X^{res}}$.}
\label{corr2pic}
\end{figure}

The last question addressed in this analysis is about the {\it correlations} of 
the lagged correlation matrices:
Are significant lagged correlations only found \emph{a posteriori} or does
the data indicate a possibility for a reasonable prediction of future lead-lag structures?
To this end we calculate the correlation of matrix elements between the lagged correlation
matrices obtained from different (non-overlapping) observation periods $T_n$ and $T_m$, 
\begin{equation}
\label{corr2formula}
\begin{split}
&c(T_n,T_m)=\\
&\frac{
\langle{(C^{ij}_\tau(T_n)-\langle{}C^{ij}_\tau(T_n)\rangle_{ij})
(C^{ij}_\tau(T_m)-\langle{}C^{ij}_\tau(T_m)\rangle_{ij})}\rangle_{ij}}
{\sigma_{T_n}\sigma_{T_m}}\quad.
\end{split}
\end{equation}
Here, the average extends over all matrix-elements and $\sigma_{T_n}$ 
denotes the standard deviation of matrix ${\bf C}_1(T_n)$.
Figure \ref{corr2pic} depicts the characteristics we obtained from empirical
data. While the expected band of correlation-coefficients would be bound by
very small
values (in the order of $1/400$), we find extremely significant correlations,
especially for the $Q=10$ case. As expected, the 'predictability' of future weighted
lead-lag matrices is significantly higher for lagged matrices calculated over longer sub-periods.
The inset of Figure \ref{corr2pic} shows $c^{\rm res}(T_n,T_m)$, i.e. the same
quantity calculated for
the residual data. Overall correlations are lower in this case, meaning
nothing else than that the market-wide movements exhibit predictable lead-lag
structures. 
However, note that for Q=10 the fluctuations
of $\mathrm{abs}(\lambda_n)$ depicted in the inset of Fig. \ref{timedep1} are not
mirrored by any specific variation of $c(T_1,T_1+d)$ in Fig. \ref{corr2pic}.

Although the present analysis of time-dependence is not comprehensive in every
respect, we may state that non-random structures prevail to quite low
information-to-noise 
ratios and that a significant amount of lagged correlation matrices
is predictable for future periods. However, shortening the length of the
sub-periods results in decreasing predictability.

\section{Conclusion}
\label{sec5}

We have applied random matrix theory to lagged cross-correlation matrices and
theoretically derived the eigenvalue spectra emanating from the respective real asymmetric
random matrices in dependence of the information to noise ratio, 
$Q$.  
Specifically, we have
shown that -- in the case of any eigenvalue 'gas' satisfying circular symmetry -- 
an inverse Abel-transform can be used to reconstruct the radial
density, $\rho(r)$, from rescaled projections available via solutions of the symmetrized
problem. 
Based on these theoretical results, we analyzed empirical cross-correlations of 5 min
returns of the S\&P500. 
For the full time-period observed, we found remarkable deviations from the 
prediction of the efficient market hypothesis and discussed various structural
properties of these deviations. We found the largest eigenvalue being
associated with a sub-matrix of exponentially distributed entries. This
eigenvalue was associated with a strong hub-like leading influence of the IT sector.
Analyzing data based on the residuals of a regression to common movements, we
found that cluster structure in the lead-lag network is strongly enhanced. 
 Looking at lagged correlation matrices pertaining to
sub-periods of the overall investigation period we found that deviations from
the theoretical prediction do occur at quite low information to noise ratios.
We also found that significant parts of the lagged correlation matrix should
be predictable via measurements of past (non-overlapping) periods.

We think that the current work can be extended in various directions.
On the theoretical side,  
a closer investigation of the nature finite-size effects in the ensemble 
of time-lagged correlation matrices and comparison with the exact finite-size 
result of the random real asymmetric case \cite{edelman} would be tempting. 
Finite-size effects could also be inferred from the terms which were found to 
vanish in the $N\rightarrow{}\infty$ limit in Appendix A.
We also think that some work is needed in an exact understanding 
of the relation between the eigenvalue spectra (including the left and right 
eigenvectors of the ensemble discussed here) and the singular value decomposition 
of related problems \cite{Bouchaud}. 
Also a rigorous study of a 'cleaning procedure' along the lines 
of methods already worked out for equal-time financial covariance matrices
could be pursued as well. 

Finally we believe that the presented work -- in
general -- should allow for an eigenvalue-dependent, systematic study 
of the influence  of matrices and their interplay with
equal time-correlations between financial assets in concrete models. 
The fact that cluster structure conforming with market sectors can be 
found in lagged correlation matrices already indicates
the direction of findings to be expected from such work.

\section*{Acknowledgements}
We thank J.D. Farmer for encouraging discussions on the matter and Jean-Phillipe 
Bouchaud for various very useful suggestions, especially for pointing out 
Eq. (\ref{expansion}) to C.B., who further acknowledges useful information from M. Biely. 
Data is by courtesy of red-stars.com data AG, the paper was sponsored in part by 
the Austrian Science Fund under FWF project P17621-G05. 

\section*{Appendix A}
Based on the series expansion (\ref{expansion}) of the potential $\phi$, 
we have calculated the first four terms in the series.
For the first term, one easily obtains
\begin{equation}
\lim_{N\rightarrow\infty}\frac{1}{N}\langle\mathrm{Tr}(B)\rangle{}_c=
\lim_{N\rightarrow\infty}\frac{1}{N}\mathrm{Tr}(\langle{}C^{ij}C^{ji}\rangle{}_c)=\frac{1}{Q}\quad.
\end{equation}
since all other terms vanish as $\mathrm{Tr}(C)$ gives just $N$ times the averages 
of the autocorrelation of the assumed iid white noise process.
For calculating the second term, it is useful
 to remember $\mathrm{Tr}(\mathbf{A}\mathbf{B})=\mathrm{Tr}(\mathbf{B}\mathbf{A})$ and
$\mathrm{Tr}({\bf C}{\bf C})=\mathrm{Tr}({\bf C}^T{\bf C}^T)$ as 
well as taking into account that odd powers of $C$ vanish. One then arrives at
\begin{equation}
\begin{split}
\lim_{N\rightarrow\infty}\frac{1}{N}\langle\mathrm{Tr}(B^2)\rangle{}_c&=
\lim_{N\rightarrow\infty}\frac{1}{N}(\mathrm{Tr}(\langle{}(C^{ij}C^{ji})^2\rangle{}_c)\\
&+(x^2+y^2)\mathrm{Tr}(\langle{}2C^{ij}C^{ji}\rangle{}_c)\\
&+(x^2-y^2)\mathrm{Tr}(\langle{}2C^{ij}C^{ij}\rangle{}_c)) \quad .
\end{split}
\end{equation}
This structure is also typical for higher order terms (not shown for brevity). 
The trace in the 'dangerous' term proportional to $x^2-y^2$ is nothing else than 
$N$ times the variance of autocorrelations which is just $1/T$ for a Gaussian process. 
Thus, in total, the term vanishes as $1/T$ in the limit $N\rightarrow\infty$ with $Q=\mathrm{const.}$, and one gets 
\begin{equation}
\lim_{N\rightarrow\infty}\frac{1}{N}{\mathrm{Tr}\langle{}(B^2)\rangle{}_c}=K+2rQ^{-1} \quad .
\end{equation}
In very similar calculations, it is easy (but tedious), to check that
\begin{equation}
\begin{split}
\frac{1}{N}\langle\mathrm{Tr}(B^3)\rangle{}=f(r)\quad\mathrm{and}\quad
\frac{1}{N}\langle\mathrm{Tr}(B^4)\rangle{}=g(r)\quad.
\end{split}
\end{equation}
The typical situation for higher order terms is similar to the one for the 
second order term, i.e. the terms in $r$ generally depend on some function 
of $Q$ and the 'dangerous' terms (like $(x^2-y^2)^2$) vanish since they remain 
constant for growing matrix size and are thus neutralized by the prefactor $1/N$.
We do not expect any different behavior for terms higher than fourth order.

\section*{Appendix B}
The uniform eigenvalue 
distribution of real asymmetric matrices in the complex plane $\mathcal{C}$ 
found in \cite{crisantisommers}
can be almost trivially recovered from Wigner's semicircle law of real 
symmetric matrices via application of the inverse Abel-transform. 
Starting from Wigner's semicircle law $\rho(\bar{\lambda})=\frac{1}{2\pi}\sqrt{4-\bar{\lambda}^2}$ 
and after proper rescaling $\rho_x(\lambda)=\frac{1}{\sqrt{2}\pi}\sqrt{4-2\lambda^2}$
we may insert into Eq. (\ref{abelsolution}) and arrive at
\begin{equation}
\begin{split}
\rho(\lambda)&=\frac{1}{\pi^2}\int_r^{\sqrt{2}}\frac{\lambda}{\sqrt{2-\lambda^2}\sqrt{y^2-\lambda^2}}\mathrm{d}\lambda\\
&=\frac{1}{\sqrt{2}\pi^2}\mathrm{arctan}\left(\frac{\sqrt{2-\lambda^2}}{\sqrt{\lambda^2-r^2}}\right)\Bigg|_r^{\sqrt{2}}=\frac{1}{2\pi} \quad .\\
\end{split}
\label{invabeltransferg}
\end{equation}
We immediately arrive at the result of an uniform eigenvalue distribution, 
\begin{equation}
\rho(r)=
\left\{
\begin{array}{cc}
\frac{1}{2\pi} & 0<r<\sqrt{2}\\
0       & \mathrm{elsewhere}
\end{array}
\right.
\quad .
\end{equation}

\section*{Appendix C}
For $Q=1$, one solution can be written in the form
\begin{equation}
 G^H_{r=1}(z)=\frac{1}{\sqrt{2}}\sqrt{1-\frac{\sqrt{z^2-4}}{z}}\quad .
\label{greennice}
\end{equation}
Note, that this equation shows a simple relation to
the resolvent of the Gaussian orthogonal ensemble ($G^S_{Q=1}=\sqrt{\frac1z G^{GOE}_{r=1}(z)}$).
The eigenvalue spectrum following from Eqs. 
(\ref{eigenvaluedensity}) and (\ref{greennice}) can then be written as
\begin{equation}
\begin{split}
\rho^{Q=1}(\bar{\lambda})=&\frac{1}{\sqrt{2}\pi}\frac{\sqrt{-\frac{1}{2}+\frac{2}{2+|\bar{\lambda}|}}}{\sqrt{\frac{|\bar{\lambda}|}{2+|\bar{\lambda}|}}} \\
=&\frac{1}{\sqrt{2}\pi}\sqrt{-\frac{1}{2}-\frac{1}{|\bar{\lambda}|}+\frac{2}{2+|\bar{\lambda}|}+\frac{4}{|\bar{\lambda}|(2+|\bar{\lambda}|)}}
\quad , 
\end{split} 
\end{equation}
and is valued on the support $\left[-2,2\right]$.
After proper rescaling and taking an expression equivalent to Eq. (\ref{abelsolution}), namely
\begin{equation}
\rho^{Q=1}(r)=-\frac{1}{\pi{}r}\frac{\mathrm{d}}{\mathrm{d}r}\int_r^\infty{}\lambda\frac{\rho_x^{Q=1}(\lambda)}{\sqrt{\lambda^2-r^2}}\mathrm{d}\lambda
\quad ,
\end{equation}
we end up with the expression
\begin{equation}
\rho^{Q=1}(r)=-\frac{1}{\pi{}^2r}\frac{\mathrm{d}}{\mathrm{d}r}\int_r^{\sqrt{2}}
\frac{\lambda\sqrt{\frac{\sqrt{2}}{\lambda}-1}}{\sqrt{\lambda^2-r^2}}
\mathrm{d}\lambda \quad , 
\end{equation}
which can be evaluated to
\begin{equation}
\begin{split}
\rho^{Q=1}(r)&=
\frac{1}{K}
\left[
2^{3/4}{}3{}r\Gamma\left(\frac{5}{4}\right)\Gamma\left(\frac{5}{4}\right)\Phi_2^1\left(\frac{1}{4},\frac{5}{4},\frac{3}{2},\frac{\lambda^2}{2}\right)
\right.\\
&\left.
-2^{1/4}\Gamma\left(-\frac{1}{4}\right)\Gamma\left(\frac{7}{4}\right)\Phi_2^1\left(-\frac{1}{4},\frac{3}{4},\frac{1}{2},\frac{\lambda^2}{2}\right)
\right]
\quad , 
\end{split}
\end{equation}
where $K=6\sqrt{\pi^5{}r^3}$, $\Gamma(x)$ denotes the Gamma-Function and $\Phi_2^1(a,b,c,x)$ is 
the hypergeometric function. It can be checked, that -- of course --
$\int_0^{\sqrt{2}}
2\pi{}r\rho(r)\mathrm{d}r=1$.

\end{document}